\documentclass[]{article}
\usepackage{xspace}
\usepackage{graphicx}
\usepackage{sidecap}

\let\jnl=\rmfamily
\def\refe@jnl#1{{\jnl#1}}%
\newcommand\aj{\refe@jnl{AJ}}%
\newcommand\actaa{\refe@jnl{Acta Astron.}}%
\newcommand\araa{\refe@jnl{ARA\&A}}%
\newcommand\apj{\refe@jnl{ApJ}}%
\newcommand\apjl{\refe@jnl{ApJ}}%
\newcommand\apjs{\refe@jnl{ApJS}}%
\newcommand\ao{\refe@jnl{Appl.~Opt.}}%
\newcommand\apss{\refe@jnl{Ap\&SS}}%
\newcommand\aap{\refe@jnl{A\&A}}%
\newcommand\aapr{\refe@jnl{A\&A~Rev.}}%
\newcommand\aaps{\refe@jnl{A\&AS}}%
\newcommand\azh{\refe@jnl{AZh}}%
\newcommand\memras{\refe@jnl{MmRAS}}%
\newcommand\mnras{\refe@jnl{MNRAS}}%
\newcommand\na{\refe@jnl{New A}}%
\newcommand\nar{\refe@jnl{New A Rev.}}%
\newcommand\pra{\refe@jnl{Phys.~Rev.~A}}%
\newcommand\prb{\refe@jnl{Phys.~Rev.~B}}%
\newcommand\prc{\refe@jnl{Phys.~Rev.~C}}%
\newcommand\prd{\refe@jnl{Phys.~Rev.~D}}%
\newcommand\pre{\refe@jnl{Phys.~Rev.~E}}%
\newcommand\prl{\refe@jnl{Phys.~Rev.~Lett.}}%
\newcommand\pasa{\refe@jnl{PASA}}%
\newcommand\pasp{\refe@jnl{PASP}}%
\newcommand\pasj{\refe@jnl{PASJ}}%
\newcommand\skytel{\refe@jnl{S\&T}}%
\newcommand\solphys{\refe@jnl{Sol.~Phys.}}%
\newcommand\sovast{\refe@jnl{Soviet~Ast.}}%
\newcommand\ssr{\refe@jnl{Space~Sci.~Rev.}}%
\newcommand\nat{\refe@jnl{Nature}}%
\newcommand\iaucirc{\refe@jnl{IAU~Circ.}}%
\newcommand\aplett{\refe@jnl{Astrophys.~Lett.}}%
\newcommand\apspr{\refe@jnl{Astrophys.~Space~Phys.~Res.}}%
\newcommand\nphysa{\refe@jnl{Nucl.~Phys.~A}}%
\newcommand\physrep{\refe@jnl{Phys.~Rep.}}%
\newcommand\procspie{\refe@jnl{Proc.~SPIE}}%
          
\newcommand{\Al}{$^{26}$Al\xspace}
\newcommand{\about}{$\simeq$}
\newcommand{\degree}{$^{\circ}$}

\newcommand{\Msol}{M\ensuremath{_\odot}\xspace}

\title{Cosmic Gamma-Ray Spectroscopy}

\author{{Roland Diehl}\\
       {\small Max Planck Institut f\"ur extraterrestrische Physik, D-85748 Garching, Germany}
        }

\begin{document}

\maketitle

\begin{abstract}
{Penetrating gamma-rays require complex instrumentation for astronomical spectroscopy measurements of gamma-rays from cosmic sources. Multiple-interaction detectors in space combined with sophisticated post-processing of detector events on ground have lead to a spectroscopy performance which is now capable to provide new astrophysical insights. Spectral signatures in the MeV regime originate from transitions in the nuclei of atoms (rather than in their electron shell). Nuclear transitions are stimulated by either radioactive decays or high-energy nuclear collisions such as with cosmic rays.  Gamma-ray lines have been detected from radioactive isotopes produced in nuclear burning inside stars and supernovae, and from energetic-particle interactions in solar flares. Radioactive-decay gamma-rays from $^{56}$Ni directly reflect the source of supernova light.  $^{44}$Ti is produced in core-collapse supernova interiors, and the paucity of corresponding $^{44}$Ti gamma-ray line sources reflects the variety of dynamical conditions herein. $^{26}$Al and $^{60}$Fe are dispersed in interstellar space from massive-star nucleosynthesis over millions of years. Gamma-rays from their decay are measured in detail by gamma-ray telescopes, astrophysical interpretations reach from massive-star interiors to dynamical processes in the interstellar medium. Nuclear de-excitation gamma-ray lines have been found in solar-flare events, and convey information about energetic-particle production in these events, and their interaction in the solar atmosphere. The annihilation of positrons leads to another type of cosmic gamma-ray source. The characteristic annihilation gamma-rays at 511 keV have been measured long ago in solar flares, and now throughout the interstellar medium of our Milky Way galaxy. But now a puzzle has appeared, as a surprising predominance of the central bulge region was determined. This requires either new positron sources or transport processes not yet known to us.  In this paper we discuss instrumentation and data processing for cosmic gamma-ray spectroscopy, and the astrophysical issues and insights from these measurements.}
\end{abstract}


\section{Cosmic gamma-rays and their spectra}
The characteristic energies of gamma-rays are measured in \emph{energy units of MeV}\footnote{One MeV, or 10$^{6}$eV), corresponds to a wavelength of about 10$^{-12}$m or a frequency of 10$^{22}$Hz.}, 5--6 orders of magnitude above the typical energies of atomic transitions which shape spectra in the optical domain. Cosmic gamma-rays thus are messengers of high-energy processes in cosmic sites. The typical binding energy of electrons in atoms are several eV, while binding energies of nucleons in atomic nuclei are of order several MeV; hence, MeV gamma-rays are often related to \emph{nuclear transitions}. Characteristic temperatures for gamma-ray emission, according to Wien's displacement law, would be 10$^9$K (='GK') for thermal gamma-rays; at such temperatures, objects would be unstable unless confined, e.g. the interior of a star may be at temperatures of millions of K as the large gravitational mass holds the object together, while nova and supernova explosions feature GK temperatures in their interiors. MeV energies are above the rest mass energy of electrons, hence electrons and positrons at these are \emph{relativistic}.   

\begin{figure}
\centering
\includegraphics[width=\textwidth]{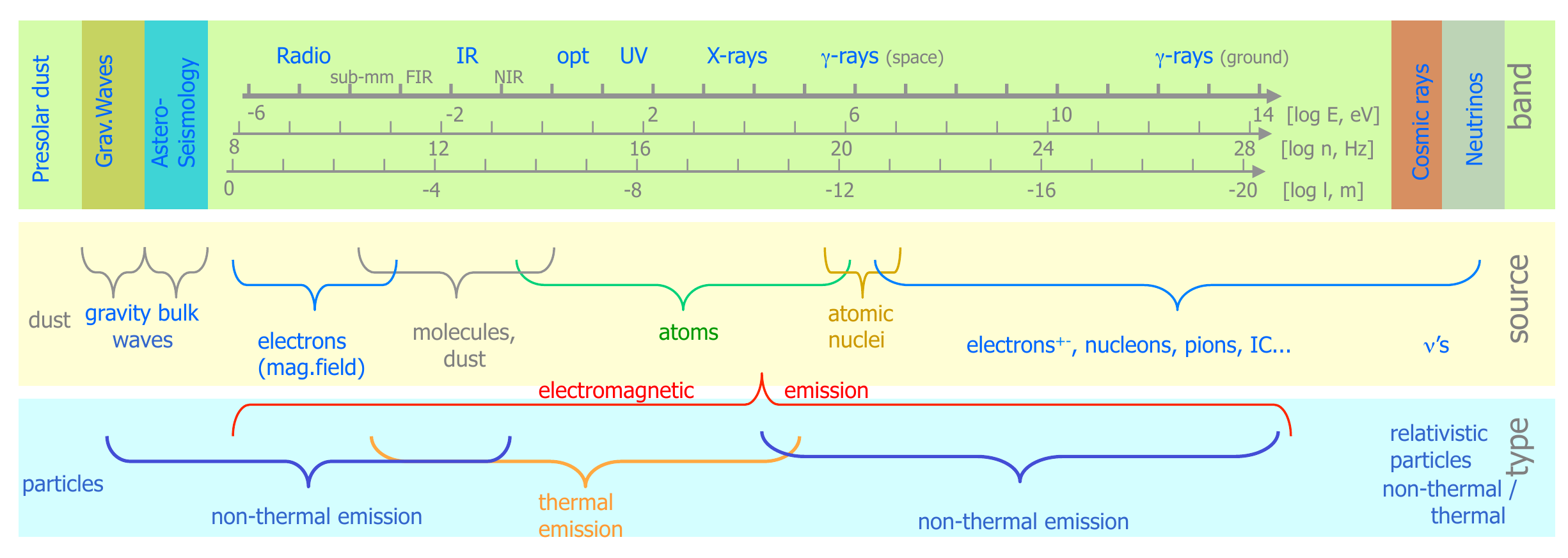}
\caption{The variety of astronomical messengers, their characteristics, and their physical information. The regime near log(E)[eV]=6 is the subject of the current paper. (Adapted from \cite{Diehl:2011}).}
\label{fig_messengers}
\end{figure}

\begin{table}[htdp]
\caption{Strongest spectral lines in the nuclear-physics part of the gamma-ray domain between \about~0.1 and 10~MeV, and their characteristics}
\begin{center}
\begin{tabular}{|c|l|l|}
\hline
energy  &  source process & astrophysical origin \\
 (MeV)     &               &   (source type)  \\
 \hline
0.078 &    radioactive decay: $^{44}$Ti               & ccSN interior nucleosynthesis  \\
0.122    &   radioactive decay: $^{57}$Ni         & supernova nucleosynthesis \\
0.478    &   radioactive decay: $^7$Be         & nova nucleosynthesis \\
0.511 &    positron annihilation           & nucleosynthesis, compact stars,  \\
  &                & dark matter   \\
0.847    &   radioactive decay: $^{56}$Ni         & supernova nucleosynthesis \\
1.157 &    radioactive decay: $^{44}$Ti               & ccSN interior nucleosynthesis  \\
1.173 &    radioactive decay: $^{60}$Fe,Co               & ccSN ejected nucleosynthesis  \\
1.275 &    radioactive decay: $^{22}$Na               & nova nucleosynthesis  \\
1.332 &    radioactive decay: $^{60}$Fe,Co               & ccSN ejected nucleosynthesis  \\
1.634 &    nuclear excitation: $^{20}$Ne               & cosmic ray / ISM interactions \\
1.809 &    radioactive decay: $^{26}$Al               & massive-star and ccSN nucleosynthesis  \\
2.230 &    neutron capture  by H             & energetic nucleon interactions  \\
2.313 &    nuclear excitation: $^{14}$N               & cosmic ray / ISM interactions \\
2.754 &    nuclear excitation: $^{24}$Mg               & cosmic ray / ISM interactions \\
4.438 &    nuclear excitation: $^{12}$C               & cosmic ray / ISM interactions \\
6.129 &    nuclear excitation: $^{16}$O               & cosmic ray / ISM interactions \\
\hline
\end{tabular}
\end{center}
\label{tab:lines}
\end{table}%

Astrophysical sources of cosmic gamma-rays, therefore, are: 

\noindent (1) nuclear-burning sites such as stellar explosions, as they release radioactive nuclides which have been produced through nuclear fusion reactions in their hot interiors \cite{Diehl:2006}. These could be supernovae, such as from core-collapse of a massive star at its terminal phase of stellar evolution (\emph{supernova types II and Ib,Ic}), thermonuclear supernovae related to disruptions of a white dwarf star (\emph{supernovae type Ia}), or novae, which are thermonuclear events on the surface of a white dwarf star, which are not disruptive to the host star itself (reviews and more detail on supernova variants can be found in \cite{Woosley:2005} and \cite{Ropke:2011}). Thermonuclear events on more-compact objects such as neutron stars and black holes will be less likely to lead to significant expansion of the nuclear-burning site, and thus remain non-transparent to gamma-rays produced herein; on neutron star surfaces, such events are termed \emph{type-I X-ray bursts} \cite{Strohmayer:2006}. Also, some products of nuclear burning in the cores of normal stars may, under special circumstances, be mixed out to the stellar surface, and hence be released as part of the stellar wind, e.g. in the Wolf-Rayet phase of stars more massive than about 25~\Msol. Several lines have been observed, from supernovae, and from radioactive nuclides accumulated in interstellar space (see Table~\ref{tab:lines}). 

\noindent (2) high-energy collisions in objects or interstellar space, which lead to excitations of nuclear levels, followed by de-excitation with accompanied characteristic gamma-ray line emission \cite{Ramaty:1979,Kiener:2012}. Such high-energy interactions also produce continuum gamma-rays through the processes of inverse-Compton scattering, Bremsstrahlung, and other radiation processes related to acceleration of charges in strong fields such as curvature radiation and synchrotron emission. For the purposes of this paper, we focus on spectral signatures rather than continuum emission. Such high-energy collisions occur in interstellar space through cosmic-ray bombardment of interstellar gas, but also near cosmic relativistic-particle accelerators such as solar flares, supernova remnants, pulsar wind nebulae, accreting binaries and supermassive black holes, and gamma-ray bursts. Often, in these sites, continuum processes dominate in brightness, and therefore only in solar flares have the characteristic spectral-line signatures from nuclear transitions been detected up to now. Interactions of low-energy cosmic rays ('LECR') appear to be closest to detection thresholds of current telescopes, while characteristic lines from nuclear transitions in accreting compact stars and their plasma jets are beyond reach at present. 

\noindent  (3) Annihilation of particles with their anti-particles, such as electron-positron annihilation, which results in a characteristic line at 511~keV energy from two-photon annihilation, and a characteristic spectral feature extending from 511 keV down into the 100~keV region, which originates from 3-photon annihilation from the triplet state of the positronium atom which is formed on the most-likely annihilation pathways \cite{Bussard:1979}. This characteristic emission has been mapped to occur in an extended region throughout the inner parts of our Galaxy, has been reported also from a few transient events in compact sources, and has been observed in great detail from solar flares \cite{Prantzos:2011}.       

\section{Telescopes for cosmic gamma-rays} 
At gamma-ray energies, interactions with the detector material of the astronomical telescope have a character different from lower energies: Up to X-ray energies, photons interact within a small volume, while at gamma-ray energies the penetration depth into any material is macroscopic, of size mm to cm. Therefore, the mirror or lens optics which characterizes astronomical telescopes at lower energies (or longer wavelengths) becomes unfeasible, and the aperture of the telescope is identical to the detector's surface area. The detector itself, correspondingly, converts the photon energy into a signal across a macroscopic depth range rather than a thin surface region. 
The physical processes which are responsible for such interaction are 

\noindent (1) Compton scattering, i.e. the inelastic scattering of the incident photon on an electron, which produces a secondary photon with reduced energy and an energized secondary electron, and 

\noindent (2) Pair production, i.e. the conversion of the electromagnetic energy of the photon into a pair of electron and anti-electron (positron) in the electric field of a nucleus of the detector material. Pair production has a threshold energy of 1.022~MeV, the rest mass of the pair of electron and positron. The excess energy and momentum of the incident photon are distributed to the pair of particles, and thus the tracks of these ionizing particles can be traced with appropriate detectors.

\begin{figure}
\centering
\includegraphics[width=0.35\textwidth]{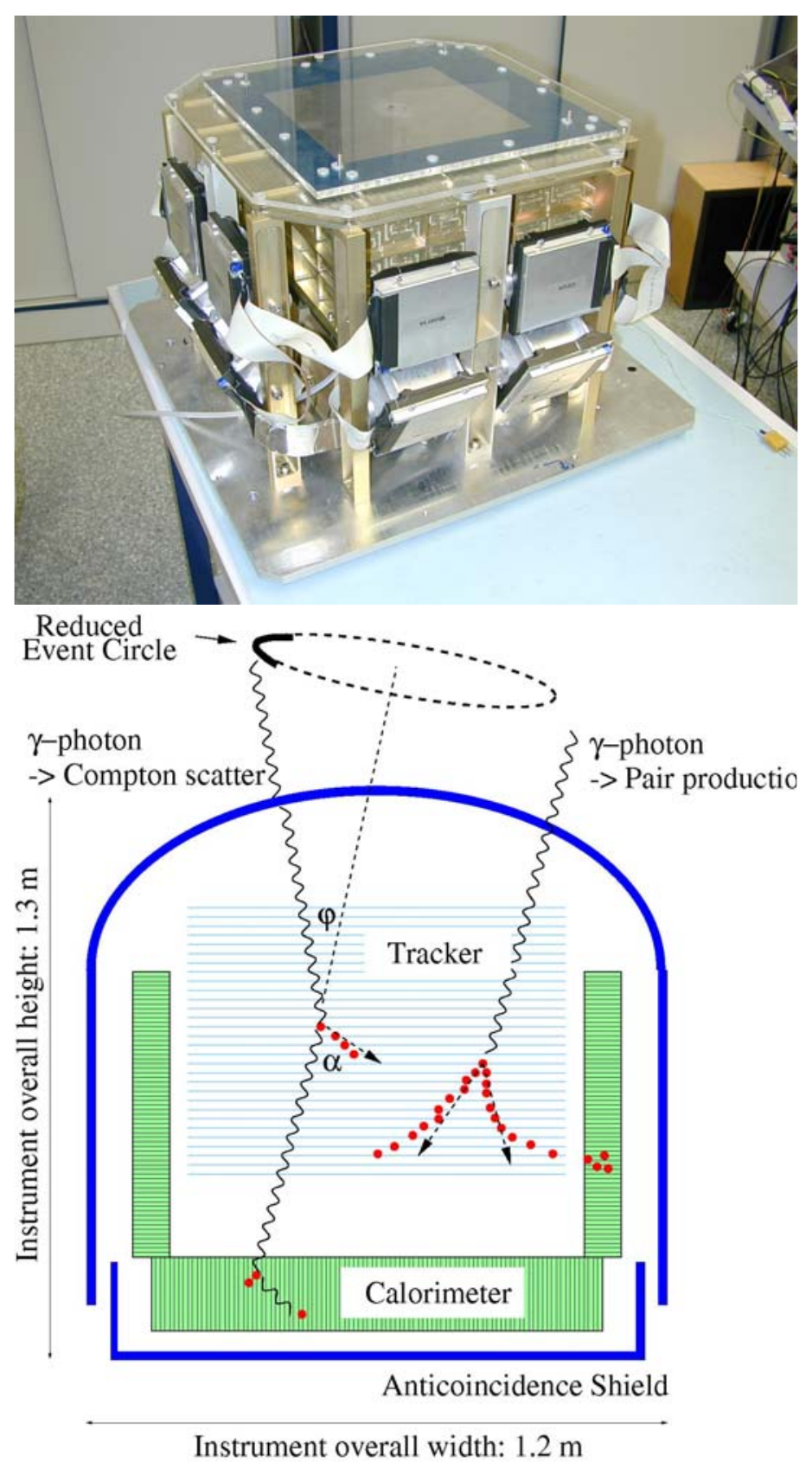}
\includegraphics[width=0.6\textwidth]{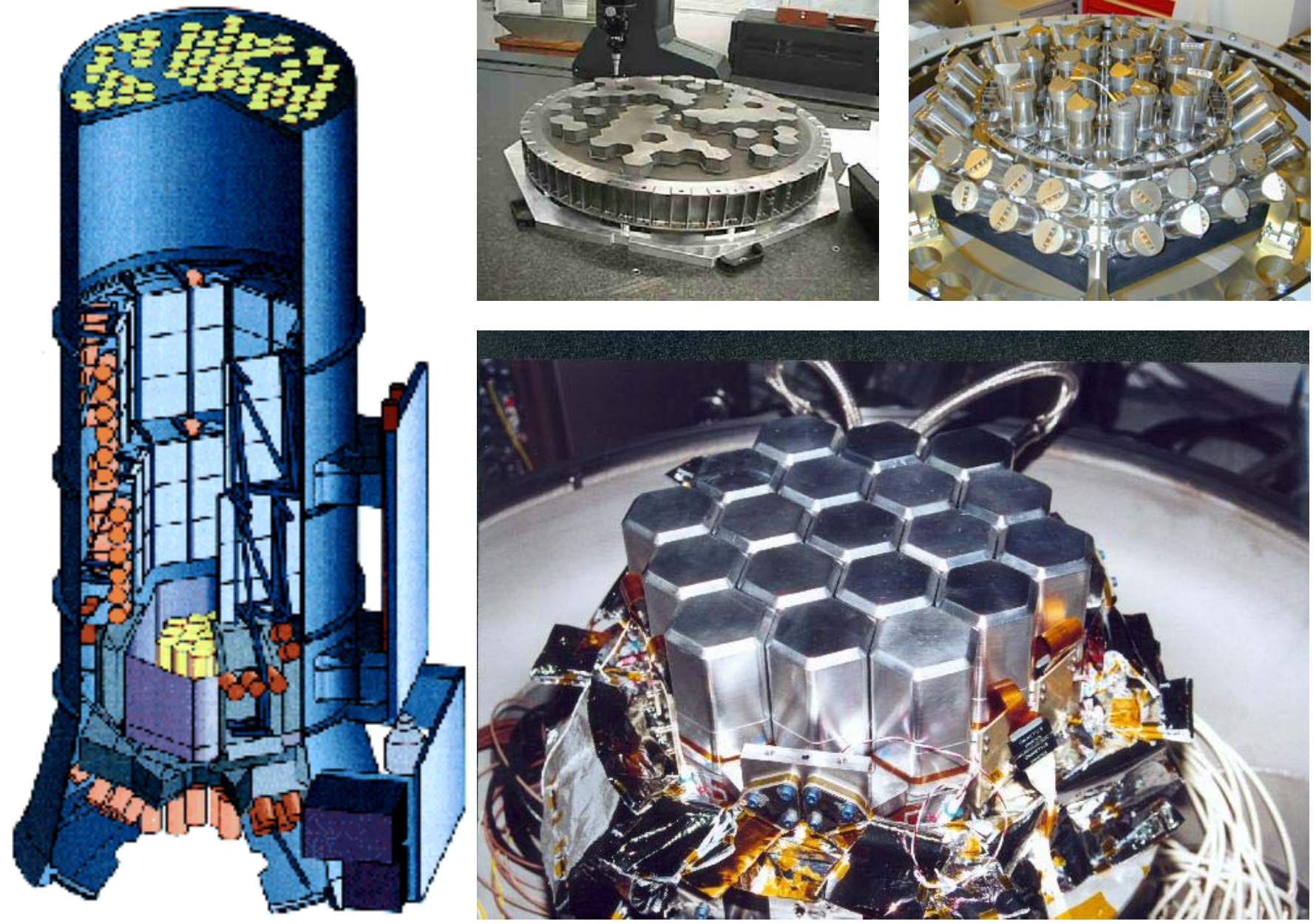}
\caption{Examples of telescopes for MeV gamma-rays. \emph{Left:} The principles of interactions of primary photons are shown in the sketch of a Compton telescope, with detector components for the primary interaction, the tracking of secondaries, the measurement of total energy, and detection of energetic charged particles causing instrumental backgrounds. The picture \emph{(on top)} shows a laboratory prototype \cite{Kanbach:2005}.
\emph{Center:} The spectrometer on INTEGRAL is a coded-mask telescope. \emph{Right:} The main instrument components, i.e. the coding mask made of tungsten and the ant coincidence detector system with 254 photomultipliers  \emph{(upper figures)}, and  \emph{(lower figure)}  the detector plane which consists of Ge detectors which optimize spectral resolution. Figure courtesy CNES. }
\label{fig_MeV-telescope}
\end{figure}

Bremsstrahlung may convert some of the particle energy of the secondary into photon energy, however, and may complicate tracking due to the larger range of photons in material compared to the charged leptons. Compton scattering results in a secondary photon, with an energy of about the same order of magnitude as the primary photon, hence with a similar range within the detector before a second interaction and partial energy deposit occurs. Therefore, successive Compton scatterings and their total range constitute the upper end of the required detector size.
The interaction depth will vary from photon to photon, reproducing the probability distribution given by the effects of Compton scattering (between about 0.1 and about 2 MeV) and pair production (at higher energies), respectively. But event by event, interaction depths will vary by macroscopic amounts. 

This leads to a degradation of imaging resolution, from the scatter of interaction depth in the detector pixel. Similarly, spectroscopic resolution depends on the homogeneity of energy deposits and its conversion into the signal amplitude across the pixel depth (and area). 
At the high-resolution limit, the intrinsic energy distribution of  the electrons in the detector material, with which the photon interacts, sets a lower limit on energy resolution: Atomic electrons which Compton-scatter the incident photon will have different energies according to their atomic orbits, and impose statistical fluctuations at the tens-eV level to the energy of the secondary electron from such Compton scatter, which together with the secondary photon carries the energy of the incident photon.  

\noindent
The secondary photons and particles which are characteristic for high-energy interaction processes of photons with materials have energies below and up to the primary photon's energy, say MeV. In contrast to detectors at GeV energies or above, the number of secondaries is rather small, and the number of total interactions from the primary interaction down to deposition of the total energy of the primary photon also is relatively small. In GeV detectors, secondaries produce copious secondary ionization, and, since energies are large, one may add thick passive intervening mass layers into the path of secondaries to simply multiply their number, thus enhancing the ionization trace of secondaries. At MeV energies, each single interactions is 'precious' and should be measured with as much precision as feasible, to maximize the information about the primary photon, specifically about its energy and its arrival direction. 

For these reasons, the technology of \emph{Compton telescopes} has been assessed to be most promising and effective as an astronomical instrument. Here, a large active detector volume ensures that both primary and secondary interactions can be measured in sufficient detail through measurements of energy depositions and momentum vectors. Still, the interpretation of these signals is not straightforward, as the Compton scattering, Bremsstrahlung, and pair processes cannot be inverted directly into reconstruction of the primary photon's energy and arrival direction \cite{von-Ballmoos:1989,Zoglauer:2008}. The statistical fluctuations inherent to measurements of energy deposits and track directions add to the resolution limitations of detector elements in spatial and energy dimensions, and are responsible for instrumental limitations.

Additionally, the lack of focusing and therefore identity of photon detection and collection areas imply that the instrumental background present in the detector itself is more important, not being suppressed by a collection/detection enhancement factor which is common to other telescopes. Moreover, instrumental background dominates the total number of measured events, and those are difficult to disentangle from the few desired cosmic-photon detection events. This is due to the physical origins of background, which are mostly the same processes that also create cosmic photons of interest: High-energy particle collisions from cosmic-ray bombardment of the instrument and its supporting structures result in continuum photons as well as nuclear line emission, as discussed above. The \emph{prompt} background events can be rejected when the primary charged particle or cosmic ray can be detected with a suitable detector system, and used in anti-coincidence to eliminate detection events which occur within a short time window after the high-energy particle traversal. But \emph{delayed} background events from short- or long-lived radioactivities or nuclear excitations cannot be rejected. These must be identified from their behavior in data space and their characteristics, as they might differ systematically from cosmic photon events. Such filtering is not perfect, due to statistical broadening of signals, and also may eliminate some desired cosmic photon events. Therefore, the processing of gamma-ray detector signals and rejection/suppression of background is complex, and often a major threshold for astronomers to make use of MeV gamma-ray telescope data in broader studies. MeV gamma-ray spectroscopy is mostly done by a small group of astrophysicists which also are specialists in such instruments.        

Cosmic gamma-ray spectroscopy was initiated by space exploration programs and nuclear radiation detectors in the 1960ies. The first cosmic lines reported were the positron annihilation line near 500~keV from the Sun and from the Galaxy, later the  detection of interstellar decay of \Al, and the first detection of supernova radioactivity gamma-rays in SN1987A. Broader studies of the gamma-ray sky were then undertaken through NASA's Compton Gamma-Ray Observatory mission (1991-2000), and through ESA's   INTErnational Gamma-Ray Astrophysics Laboratory (INTEGRAL) mission. Gamma-rays from the Sun and its flares had been studied with the Solar Maximum Mission (SMM) (1980-1989) and its scintillator-based Gamma-Ray Spectrometer instrument (Forrest et al. 1980), and later with the Reuven Ramaty High-Energy Solar Spectroscopic Imager (RHESSI; 2002--present) \cite{Lin:2002,Smith:2003}, which featured Ge detectors and corresponding high (keV-type) spectral resolution at gamma-ray energies. Several balloon-borne instruments contributed also significant spectroscopy results, such as the Ge detector equipped experiments of GRIS \cite{Teegarden:1985} and Hexagone \cite{Durouchoux:1993a} in the 90ies, and  more recently the TGRS experiment on the WIND spacecraft \cite{Seifert:1996} and the Nuclear Compton Telescope  (NCT) balloon experiment \cite{Boggs:2011}. 

CGRO had two gamma-ray spectroscopy instruments on board, the COMPTEL Compton telescope \cite{Schoenfelder:1993a}, and the OSSE Spectrometer \cite{Johnson:1993}; both featured a modest spectral resolution of order 10\% (FWHM). This is inadequate for any gamma-ray spectroscopy in the sense of identifying new lines or constraining kinematics of source regions through line shape measurements, as we now know from high-resolution instruments based on semiconductor detectors, such as INTEGRAL's SPI spectrometer which provide resolutions of \about~0.1\% . The GRSE instrument which was originally foreseen to address nuclear-line spectroscopy on this spacecraft, was removed during mission preparation following cost overruns. A 'Nuclear Astrophysics Explorer (NAE)' \cite{Matteson:1991a} mission was discussed for a while, and was integrated later through a gamma-ray spectrometer into ESA's INTEGRAL mission concept \cite{Winkler:2003}. But the first-generation instruments on CGRO were essential in exploration of the gamma-ray sky, and COMPTEL's imaging capabilities at MeV energies are still the best we have.

The INTEGRAL space mission of ESA \cite{Winkler:2003} is a current example of cosmic gamma-ray instrumentation. The satellite platform hosts two main instruments employing the coded-mask technique of imaging, the SPI spectrometer \cite{2003A&A...411L..63V} and the IBIS imager \cite{2003A&A...411L.131U}, in addition to monitor instruments, the JEM-X coded-mask X-ray telescope addressing lower energies with a wider field of view, the OMC optical monitor camera for simultaneous optical exposure of the target sky regions, and the IREM charged-particle monitor detector system recording the cosmic-ray irradiation. 
The field of view of the main instruments are of similar size of order 10\degree  (IBIS 9x9\degree, SPI 16x16\degree, fully-coded parts, corner-to-corner). The large field of view  is defined by the hexagonal arrangements of the 19-element Ge detector camera (70~mm deep, 55~mm wide detectors, covering a densely-packed 268~mm wide camera plane) and the 127-element tungsten mask with 6~cm wide elements, which is placed 170~cm above the camera itself. 
 \emph{Dithering} the satellite pointing direction around a celestial target direction in \about~2\degree steps helps to add more modulation to the sky signal, so it can be separated from the (\about~constant) instrumental background. 
The background should not vary within small attitude changes, and a very eccentric 3-day orbital period reaching far out and away from radiation belts (maximum distance from Earth 150,000~km)was chosen to provide slow background variations.
Typically, the satellite pointing is changed every \about~1800~s, and observations way vary from tens of ks to Ms, depending on science target. 

The SPI spectrometer is INTEGRAL's instrument for gamma-ray spectroscopy. It features a set of 19 coaxial Ge detectors operating in the energy range 15--8000~keV, with a  detection area of 250~cm$^2$ total and a nominal sensitivity of \about~3~10$^{-5}$~ph~cm$^{-2}$s$^{-1}$ (3$\sigma$, 1~Ms observing time.  SPI's  energy resolution is typically E/$\delta$E=500 or 3~keV at 1300--1800~keV, due to high-purity n-type Ge semiconductor detectors operated at \about~80~K and \about~4~kV, with annealings maintaining this spectral performance \cite{2003A&A...411L..91R}. This allows spectroscopy of nuclear lines through unambiguous identification of line energies. 

Other current facilities in the field of cosmic gamma-ray spectroscopy include the RHESSI solar observatory \cite{Lin:2002}, the Gamma-Ray Burst Monitor scintillation detectors (modest energy resolution (10\% FWHM)) aboard the \emph{Fermi} gamma-ray satellite launched in June 2008 \cite{Meegan:2009}, and the NuSTAR mission \cite{Harrison:2013} launched in June 2012 and equipped with a focusing hard X-ray mirror and detectors operating between 6 and 79 keV, thus including lines from radioactive decay of $^{44}$Ti at 68 and 78~keV,  the lowest-energy lines known from any cosmic radioactive decay. 


\begin{figure}
\includegraphics[width=0.55\textwidth]{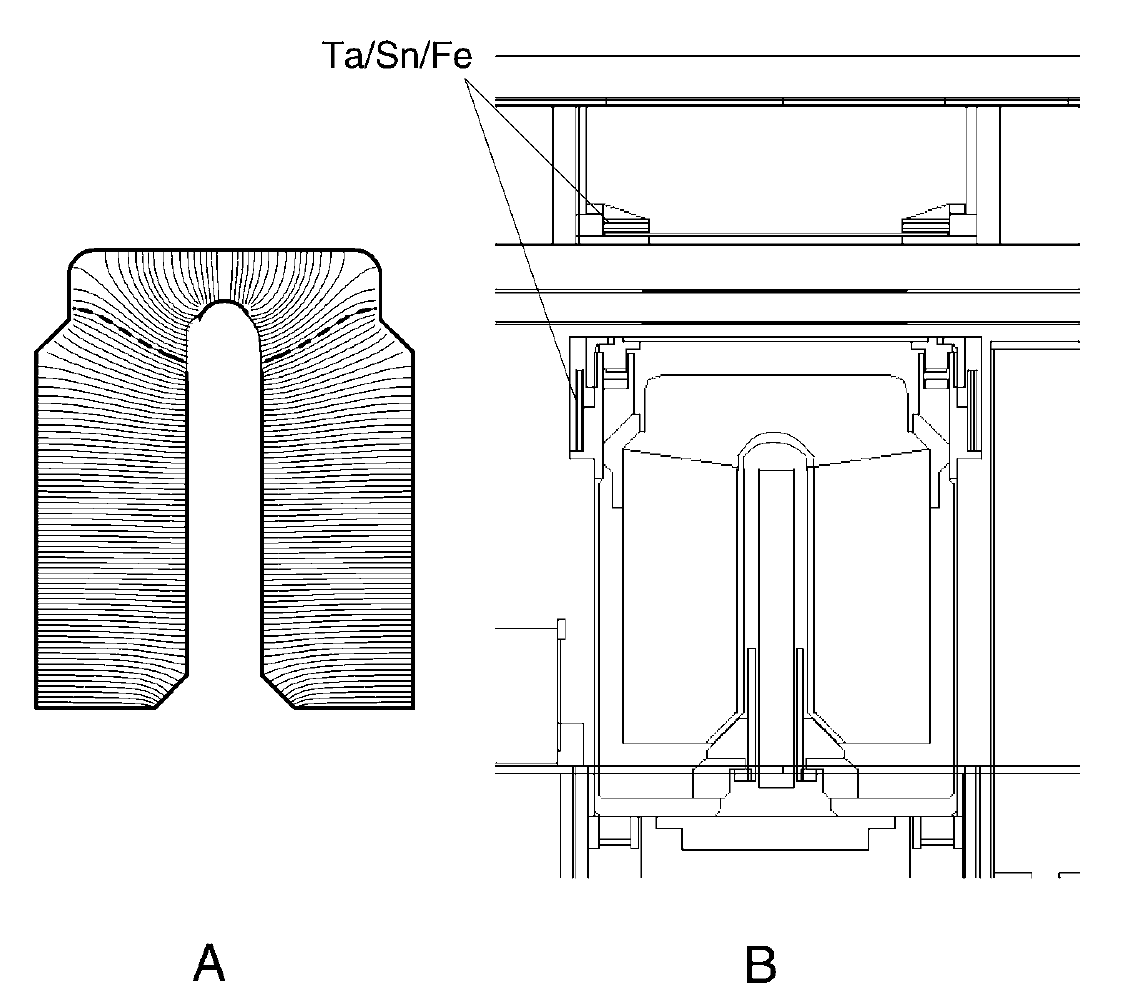}
\includegraphics[width=0.4\textwidth]{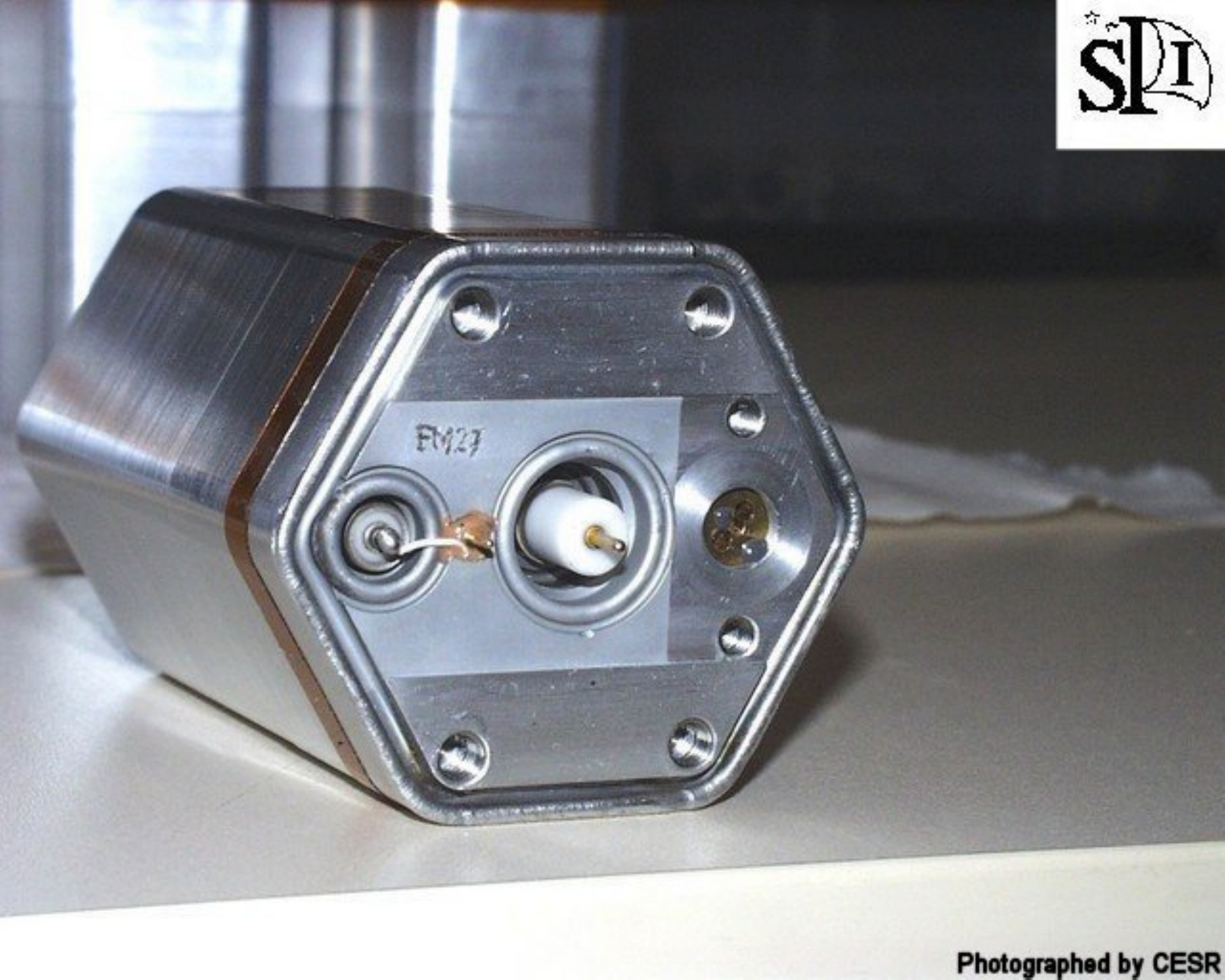}
\caption{The Ge detector for the RHESSI instrument \emph{(Fig. B, center)}. Charge collection across the detector volume is somewhat inhomogeneous due to the electric potential inhomogeneities \emph{(Fig. A left)}, as the electrode geometry implies. The photo \emph{on the right} shows a detector of SPI on INTEGRAL (photo courtesy CNES).}
\label{fig:GeDet}
\end{figure}

\section{Gamma-ray spectrometry}
Detectors for gamma-ray spectrometry in the 0.1~MeV to GeV range could be based on scintillators, solid-state detectors, drift or time projection chambers, and trackers made up of spark chambers or solid state detector stacks. The physical challenge is to produce a cascade of inelastic interactions of the primary photon and its secondaries within the volume of the detector, and to produce an electrical signal which is proportional to the total energy deposit of the cascade. High levels of background radiation lead to detectors which respond quickly, and have short dead times. 

Here, scintillation detectors have an advantage. The issue with scintillation detectors is to ensure a homogeneous and linear light collection over the volume of the scintillation detector. Imperfections result in different signal amplitudes per energy deposit, depending on the location of interactions within the detector volume. The spectral resolution required for identification of gamma-ray lines and relating them to specific nuclear transitions practically can only be achieved through solid state detectors.  

Solid state detectors operate through collection of the charge liberated from photon interactions as electrons are activated into the conduction band. In semiconductor detectors, a small band gap of few eV only allows very sensitive high-resolution detectors. Germanium detectors have been established as standard in terrestrial nuclear-physics experiments, and also space borne cosmic gamma-ray experiments; Ge detectors have been reviewed recently \cite{Vetter:2007}. In recent years, CdZnTl detectors have become popular, because they can be operated at room temperature, rather than the cryogenic temperatures required for Ge detectors, at nearly similar performance. 

\begin{figure}
\centering
\includegraphics[width=0.98\textwidth]{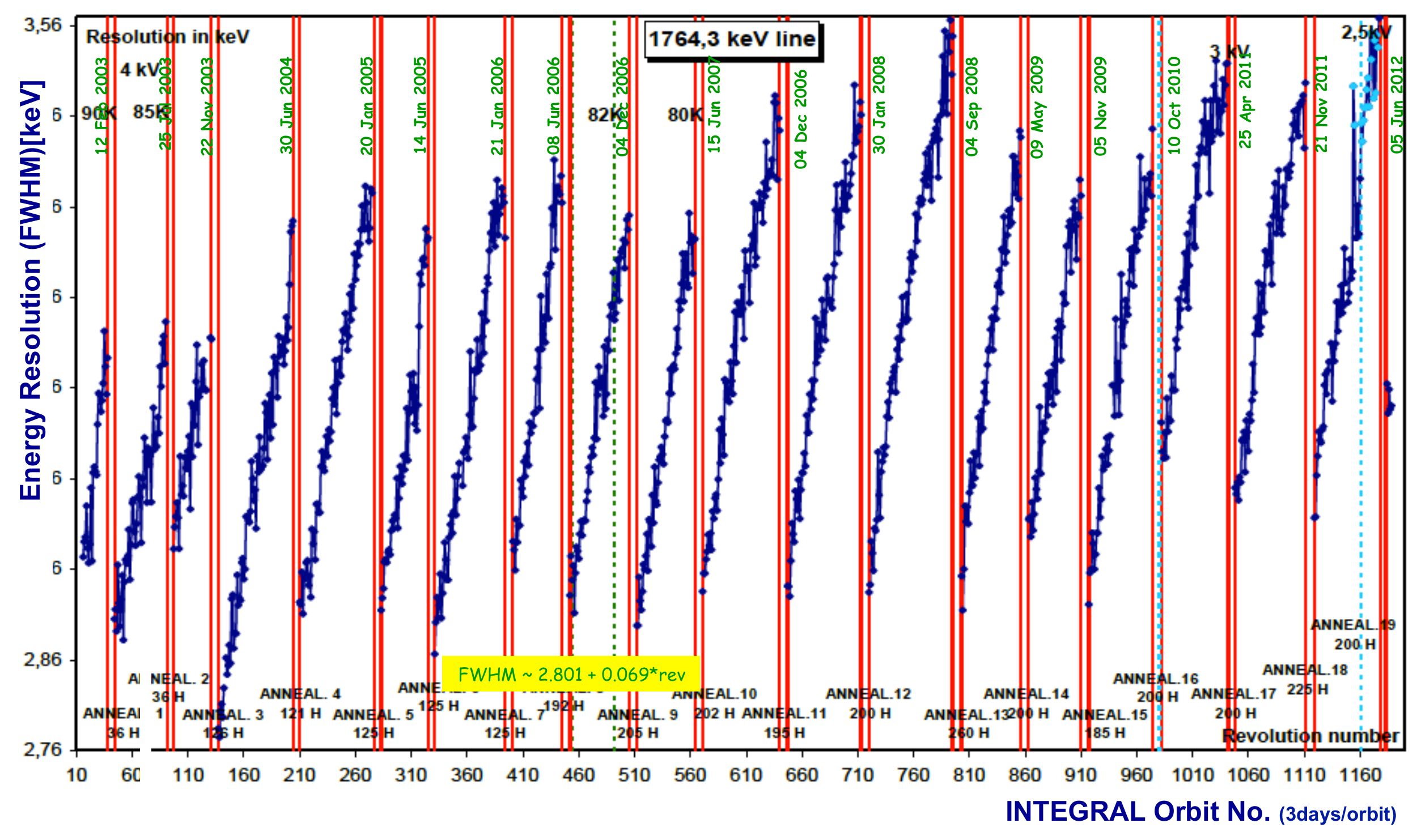}
\caption{The time history of spectral resolution of Ge detectors of the SPI instrument on INTEGRAL. Detector charge collection degrades due to cosmic-ray bombardment. Annealing operations every few months successfully restored the original energy resolution.}
\label{fig:SPI_Annealings}
\end{figure}

Charge collection implies propagation of electrons and 'holes' within the detector volume, which is achieved through an electric field between the two electrodes of the detector. Depending on detector and electrode geometry, charge collection also may be inhomogeneous across the detector volume, thus limiting spectral resolution. Moreover, bombardment with cosmic rays in the typical space environment where such detectors for cosmic photons are operated will destroy the crystal properties locally within the detector volume, thus producing traps for charge transport and therefore impeding charge collection with cumulative cosmic-ray exposure. Figure \ref{fig:SPI_Annealings} demonstrates this effect for the SPI detectors on INTEGRAL, showing the resolution for an instrumental line at 1764~keV as a function of time (in units of satellite orbits called 'revolutions' which last 3 days each) over ten years. The degradations are repaired through annealings every few months: Here, the entire detector platform is heated from cryogenic operational temperatures around 80~K to about 370K for 10-14 days, thus curing the crystal defects which impeded charge collection.

\begin{SCfigure}
\centering
\includegraphics[width=0.7\textwidth]{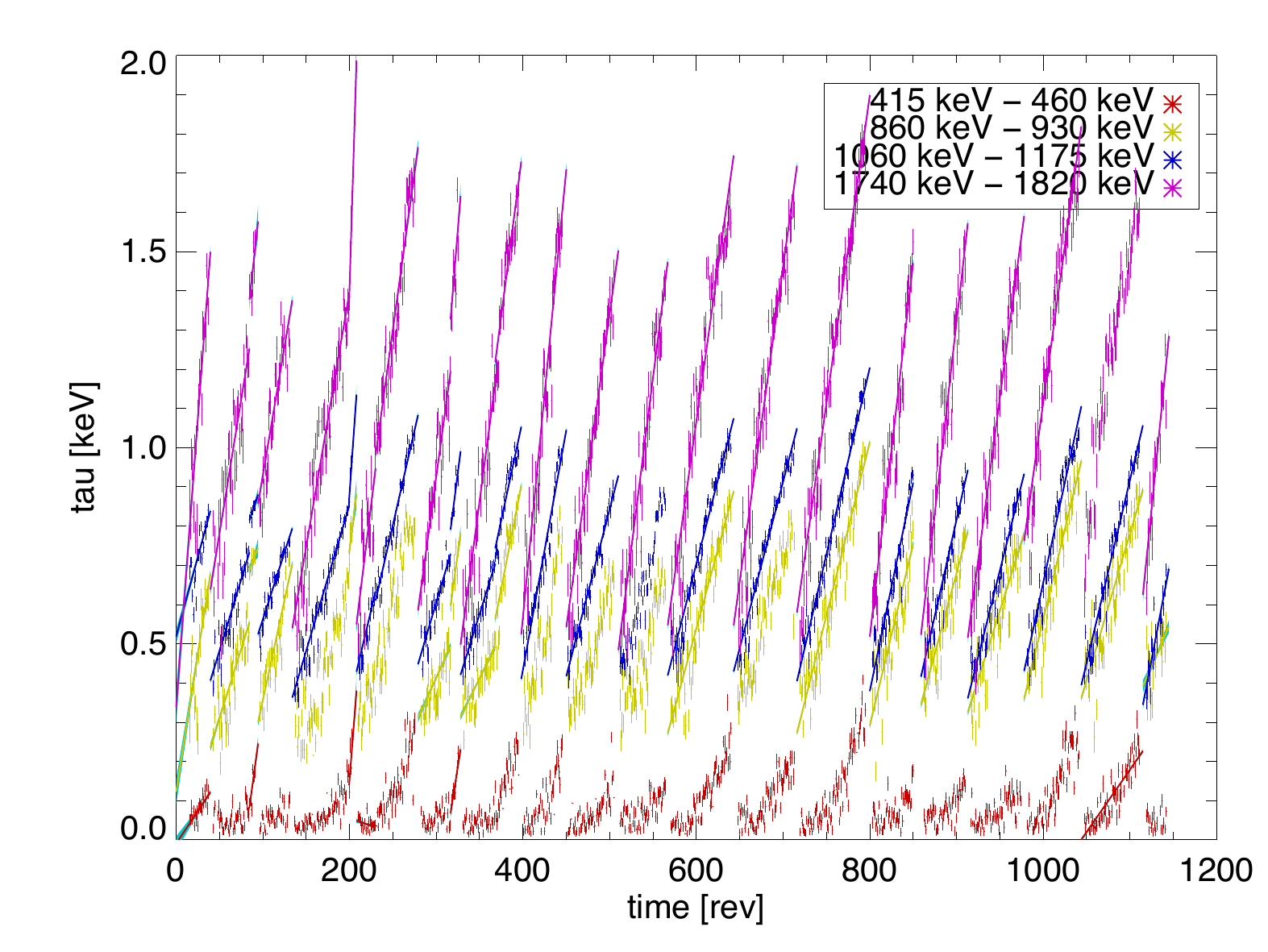}
\caption{The time history of the degradation component of the SPI instrument's energy response, as derived in four different energy bands.}
\label{fig:SPI_degradation}
\end{SCfigure}

The spectral response of the Ge detectors on SPI is monitored and stored in precise detail, from many instrumental lines fitted by a parametrized representation of the response to photons of specific energy.  The response function is composed of a Gaussian centered at the incoming photon energy, a step function from this energy towards lower energies, and a one-sided exponential from the incoming photon energy downward to represent the degrading charge collection efficiency. Figure \ref{fig:SPI_degradation} shows how the degradation thus is extracted in consistent and precise detail.  From a set of instrumental-background lines, the spectral response is decomposed into a nearly-Gaussian part plus a one-sided exponential that describes degradation of charge collection. This response is then used in spectroscopic analysis later on.

Spectroscopy at MeV energies suffers from a dominance of instrumental background in measured events, the cosmic photons of interest contribute only a small ratio of events at the percentage level or below. Typical source intensities in lines are 10$^{-3}$~ph~cm$^{-2}$s$^{-1}$ for the strongest line at 511 keV to 10$^{-6}$~ph~cm$^{-2}$s$^{-1}$ for the weakest lines detected so far (from $^{60}$Fe decay), and many predicted lines from objects of interest are about 1-2 orders of magnitude fainter still. Thus, a critical component of cosmic gamma-ray spectroscopy is the determination of the instrumental background as it may vary across the months of typical exposure times needed to collect the cosmic photons from sources of interest. The physical processes which cause the instrumental background are rather clear: Cosmic ray and secondary energetic-particle interactions with the spacecraft and instrument materials causing inelastic high-energy photon production and nuclear excitations. Nevertheless, the prediction of instrumental backgrounds from software implementations of such physical processes (e.g. GEANT) fail to achieve the required precision of better than a \%-range precision, with their typical uncertainties of 30\%. These uncertainties derive from unknown energetic-particle spectra and flux variations, and from limitations in modeling the spacecraft and instrument in its geometry and materials composition at sufficient precision; small admixtures of specific isotopes are often responsible for the strongest background lines. Moreover, those energetic particle interactions also create radioactive isotopes within the spacecraft which are not part of the materials library of these software packages, as they have been established from terrestrial high-energy physics experiments. Therefore, the method of choice for background determination is empirical fitting of models which describe the background in its spectral and temporal variations. The main challenge is to find a background representation which is orthogonal, or rather least degenerate, with variations of the cosmic photon signal. 

\begin{SCfigure}
\includegraphics[width=0.7\textwidth]{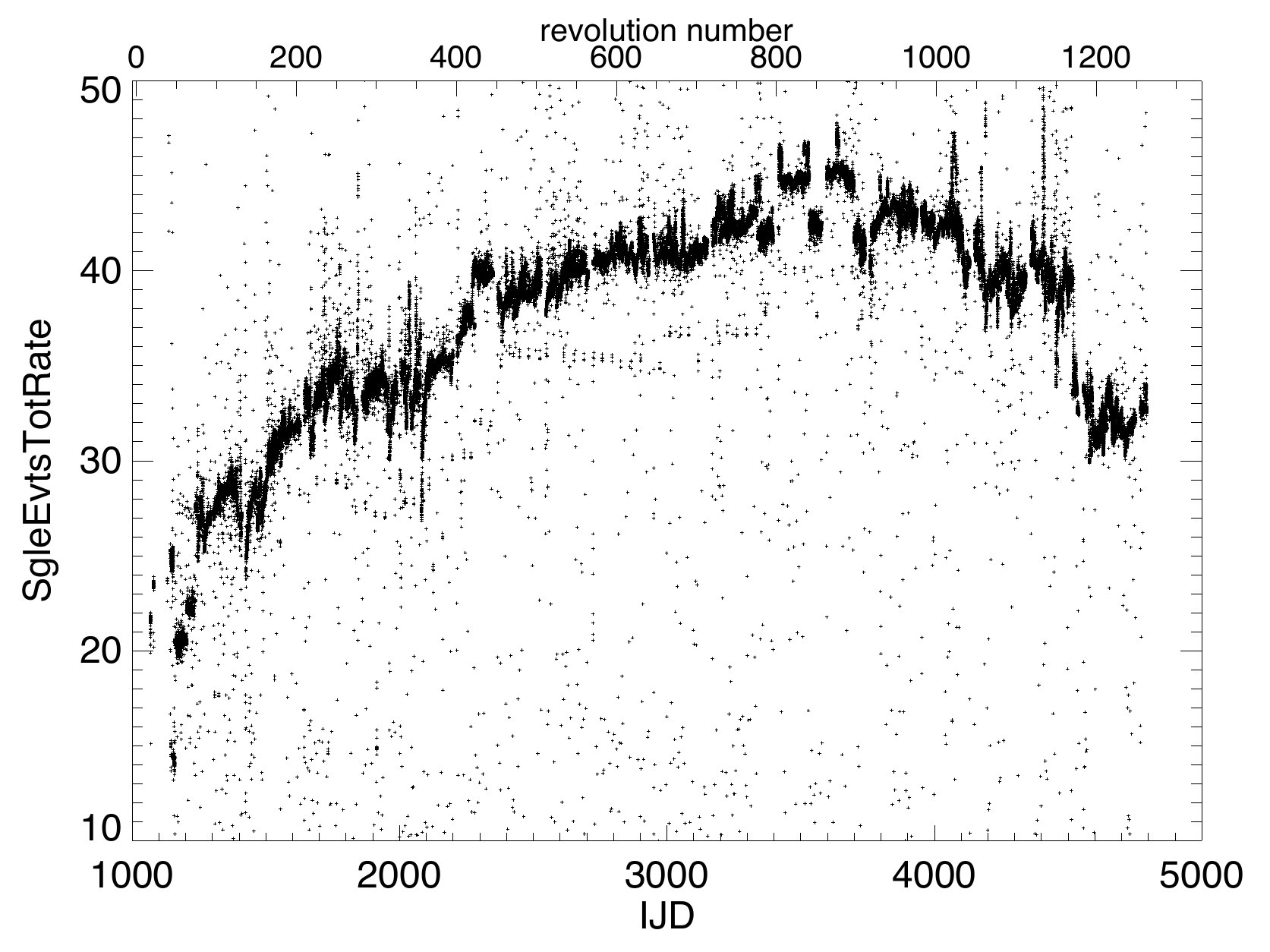}
\caption{The time history of the total SPI Ge detector event rate over ten years of the INTEGRAL mission. The lower x-axis shows time in day units (internal/truncated Julian days) while on the upper x-axis units are the INTEGRAL orbits. Solar minimum was near IJD 3400.}
\label{fig:SPI_eventRate}
\end{SCfigure}

\begin{SCfigure}
\includegraphics[width=0.7\textwidth]{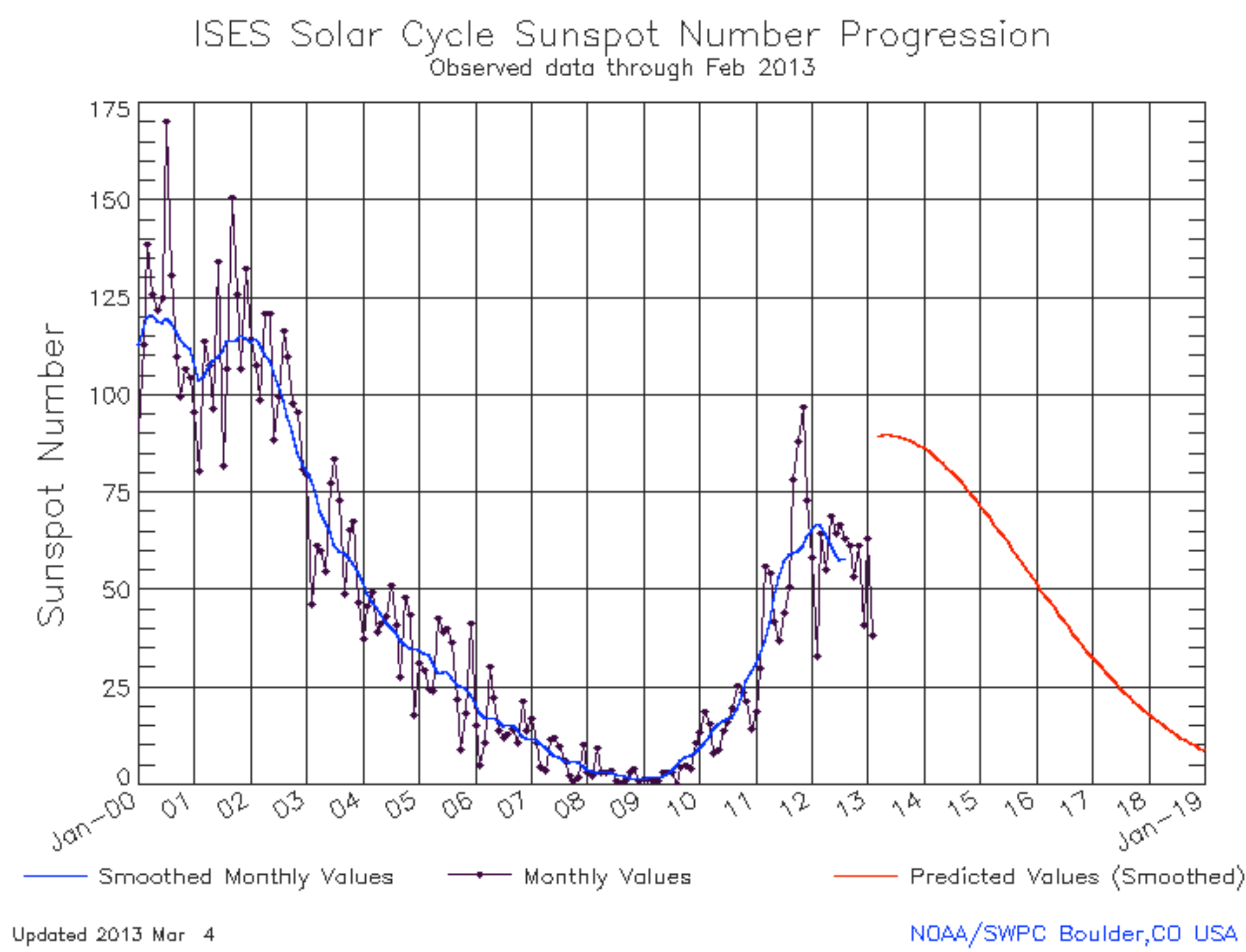}
\caption{The time history of Sunspot counts over recent years. (from space weather predictions of the US National Oceanic and Atmospheric Administration (NOAA), website http://www.swpc.noaa.gov.}
\label{fig:Sunspots}
\end{SCfigure}

Figure \ref{fig:SPI_eventRate} shows the count rate variation over ten years of the SPI detectors on INTEGRAL. The general increase of background over the first years corresponds to a decline of solar activity, as energetic particles from the Sun (solar wind) change the morphology of the Earth's magnetosphere. In near-Earth orbits of space satellites, therefore in phases of high solar activity the shielding of cosmic rays is more effective. This can be seen in Fig.~\ref{fig:SPI_eventRate} from background declining again as the solar activity picks up in cycle \#24 after the year 2010 (see Fig.~\ref{fig:Sunspots}). Superimposed are short-term rate variations, which arise from solar flare events, or magnetic storm particle events which expose the satellite to an increased particle flux for a short time. Additionally, the approach of radiation belts at the perigee of the satellite orbit leads to strongly-increased energetic particle irradiations; but here the instrument is turned off to avoid damages from saturations, so that only a declining activation signature from this perigee exposure can be seen after the instrument is turned on again leaving the radiation belt zones. The INTEGRAL orbit has been chosen to have a high eccentricity with an apogee of about 150,000~km, so that the few minutes of post-perigee activation decline are rather negligible and background generally is stable across days of this 3-day orbit.

For modeling of instrumental backgrounds, it had turned out to be most effective to make use of other, independent, detectors of energetic-particle-flux related events near the cosmic photon detectors. In the case of SPI, the rate of saturated events in Ge detectors has proven valuable, as has the rate of charged-particle detections in the anticoincidence shield detector (based on BGO scintillator material) and a veto detector plate below the coded mask (which is based on a plastic scintillator) \cite{Vedrenne:2003}. Moreover, Ge detector counts in continuum energy bands adjacent to the line of interest proved valuable. But all of these \emph{tracers} of instrumental background have been found to change their functional relation to background on time scales of days, thus require scaling parameters to be fitted during the cosmic source exposures. Therefore, a typical spectroscopy study with SPI consists of fitting a multi-parameter model to observed spectral counts per the Ge detector set and thousands of 30-minute pointings of overlapping sky regions. The model includes instrumental background through a few tracer components adjusted at different time intervals, and a spatial-distribution model of the cosmic photon sources of interest, which is folded through the instrument's imaging response to predict the cosmic event contribution. Once the intensity scaling of the cosmic source is determined in fine energy bins across the line of interest, the spectral response is then accounted for to determine the line position and shape parameters for the cosmic source. An illustration of this procedure can be found below, from the detailed spectroscopy of the celestial \Al line.

\section{Lessons in cosmic gamma-ray spectroscopy}
\subsection{Supernovae of Type Ia}\label{sources_snia}
Understanding the progenitor systems and the dynamics of the Carbon-burning triggered explosion of supernovae of Type Ia is motivated by two main reasons. SNIa are the dominating sources of cosmic iron and similar elements, and therefore key to the metal enrichment of the universe. SNIa also are key to current cosmology studies, being used as standardizable sources with known intrinsic brightness for tracing cosmological expansion \cite{2011ARNPS..61..251G}. But up to now, no consistent physical explanation of SNIa explosions has been established, in spite of many proposed scenarios and progenitors. Currently-discussed models are  (i) explosions triggered in white dwarf central regions as the Chandrasekhar mass limit is reached, (ii) explosions triggered by an off-center event such as a He-shell flash on white dwarfs of a broader mass range, and (iii) explosions resulting from merging of two white dwarfs as it may terminate the evolution of a binary system. More complex scenarios of binary evolution with recurrent pulsations culminating in a supernova are also discussed \cite{2011PrPNP..66..309R}. In all cases, the runaway nuclear carbon burning will provide the energy which disrupts the white dwarf; this is the commonly agreed part of SNIa models (see\cite{2011LNP...812..233I,2011PrPNP..66..309R} for recent reviews). 

The issues discussed relate to the nuclear C-burning flame propagation in the different density regimes, between the core of the white dwarf and the exploding outer layers.   As a result, the SNIa nuclear ashes may range from almost-pure (near 1~\Msol) $^{56}$Ni as it would result from complete burning in Nuclear Statistical Equilibrium (NSE), through deflagration with NSE and incomplete-NSE burning,  down to $^{56}$Ni amounts below 0.1~\Msol and mostly intermediate-mass element ashes resulting from partial burning , such as expected for double-degenerate merger or detonation models \cite{2007Sci...315..825M}. Correspondingly, the spatial distribution of $^{56}$Ni ashes may range from central-only  to distribution throughout the entire exploding supernova.  
The initial brightness of SNIa is due to the sudden deposition of the explosion energy of  \about~10$^{51}$erg. But soon, after a few days, supernova brightness dominated by the energy deposition of gamma-rays and positrons from radioactive decay of $^{56}$Ni. Clearly, the brightness critically depends on how $^{56}$Ni deposits its radioactive energy into the gas of the rapidly-expanding envelope, leading to thermal emission of the envelope gas at its (inward-moving) photosphere. A (bolometric, thermal-emission) luminosity maximum is obtained at \about~20 days after explosion in optical emission. Radiation propagation models are used to relate this low-energy photon emission to the original $^{56}$Ni and its radioactive energy deposition (\about 3~MeV per decay of a $^{56}$Ni nucleus) \cite{2007ApJ...662..487W}. Measurements of the total $^{56}$Ni content  through gamma-rays are more direct, and rather independent (at times > 100 days) of the density structure of the expanding supernova. The expanding supernova becomes transparent to gamma-rays on a time scale of \about~100~days, so that $^{56}$Ni decay energy deposition fades. The rise of radioactive gamma-ray luminosity, as well as gamma-ray line shapes and line-to-continuum ratio convey information about the large-scale spatial distribution of $^{56}$Ni and the envelope structure. Typically, the gamma-ray luminosity maximum is expected at 70--90~days after explosion, much later than the luminosity maximum in thermal (optical to infrared emission) supernova light. The apparent optical luminosity fading with a \about~100-day time constant thus is fortuitous, and results from radioactive decay of $^{56}$Ni through $^{56}$Co to $^{56}$Fe convolved with energy deposition properties of the rapidly-expanding supernova envelope. 

Sufficiently-nearby events, closer than about 5~Mpc, should occur once every 2--3 years (as estimated by \cite{2008NewAR..52..377I} from different SN rate proxies in the local universe), although numbers vary by factors \about~2 depending on the assumptions about the supernova rate in this more-heterogeneous population of nearby galaxies.  

The CGRO mission in 1991--2000 had two  opportunities: SN1991T occurred early in the mission, at a distance of 13 Mpc the detected hint for the characteristic gamma-ray lines appeared surprising. But it was quickly recognized that SN1991T was an anomalously-bright event also from other observed phenomena, and hence not typical for SNIa. The second event during the CGRO mission was SN1998bu at \about~17~Mpc. It was much fainter, and no gamma-ray emission could be detected \cite{2002A&A...394..517G}. 

\begin{SCfigure}
\centering
\includegraphics[width=0.65\textwidth]{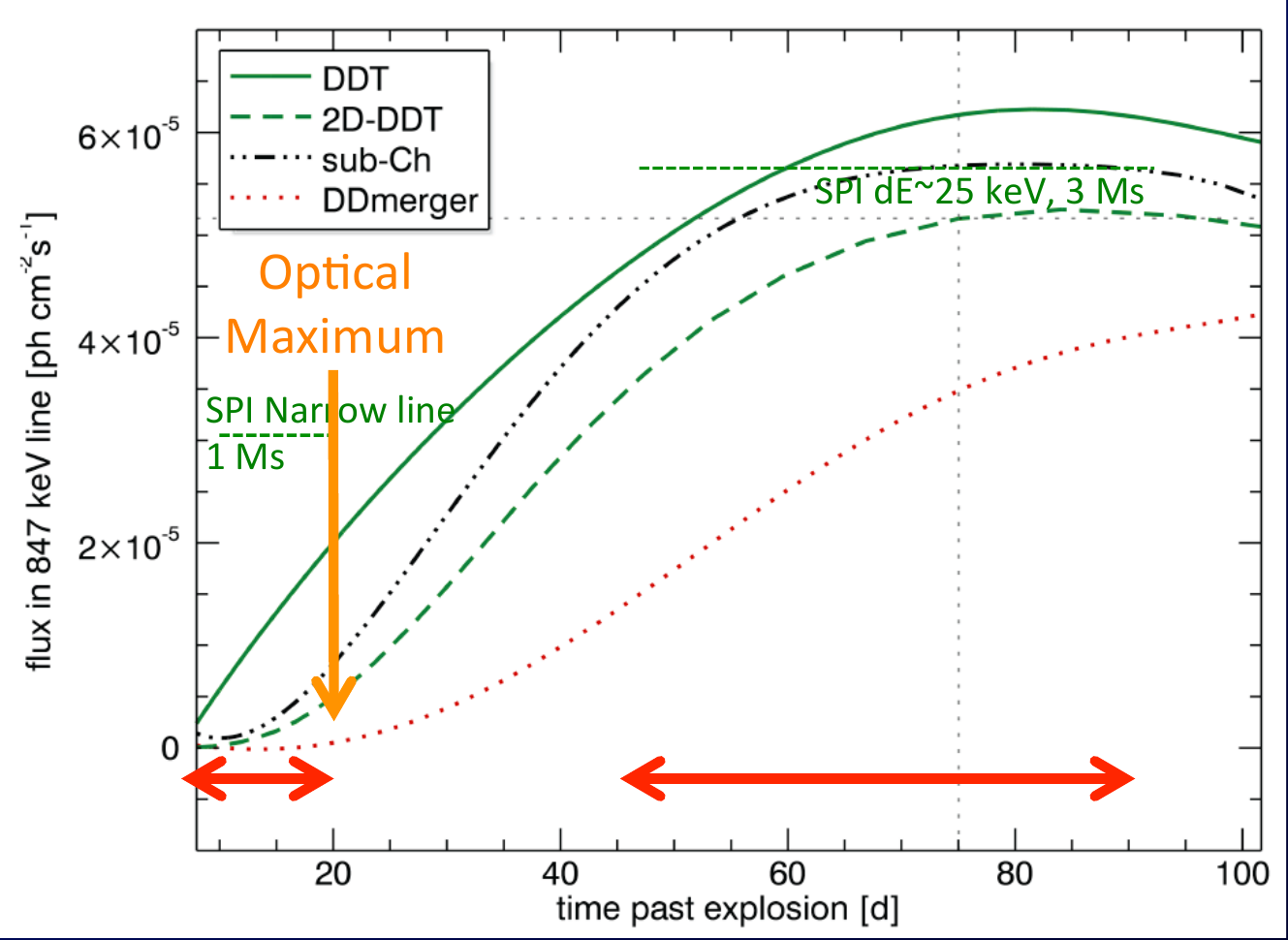}
\caption{The gamma-ray brightness evolution of a SNIa for different explosion models. Maximum brightness at optical wavelengths occurs about 20 days after the explosion (orange arrow), while gamma-rays from the $^{56}$Ni decay chain evolve much slower and differently for different  models. For the case of SN2011fe shown here, INTEGRAL observing times and SPI instrumental sensitivities are indicated (red arrows, horizontal lines). }
\label{fig_SN2011fe}
\end{SCfigure}

During the INTEGRAL mission,  two SNIa were close enough for useful INTEGRAL observations, but no signal could be found. SN2003gs probably was much too distant, at 16~Mpc. 
SN2011fe occurred on August 24, 2010, in the nearby galaxy M101/NGC5457 \cite{2011Natur.480..344N}, at a distance of 6.4 Mpc. With lines likely to be significantly broadened, the high spectral resolution does not help, considering the instrumental-background level. INTEGRAL observed SN2011fe for 4~Ms total, only to provide upper limits \cite{2011ATel.3683....1I} (see Fig.~\ref{fig_SN2011fe}). 

Thus, after the COMPTEL and INTEGRAL gamma-ray telescopes operating for 20 years, SNIa gamma-rays still have not contributed to help understand SNIa. A significant advance in sensitivity is needed to not depend only on luck with sufficiently nearby events, to below 10$^{-6}$~ph~cm$^{-2}$s$^{-1}$ as shown in several proposed mission concepts (see ESA's Cosmic-Vision program selection, e.g. \cite{Greiner:2012,von-Ballmoos:2012,Lebrun:2010}). 

\subsection{Core-Collapse Supernovae}\label{sources_ccsn}
Massive stars with initial masses above 8--10~\Msol evolve through a series of central nuclear fusion stages beyond hydrogen burning up to silicon burning.  As this energy reservoir of  the release of nuclear binding energy is exhausted eventually, gravitational collapse will terminate this evolution after 10--100~My \cite{2007PhR...442...38J,2011LNP...812..153T}. The collapse occurs on a time scale of seconds, once either central nuclear burning starves from exhaustion of the fuel for energy-liberating fusion reactions or electron capture sets in and takes away the pressure support from the (degenerated) electron gas. Infalling nuclei are decomposed in the shock wave above the newly-forming neutron star into nucleons and $\alpha$~particles, which consumes part of the gravitational energy, but neutrinos emitted by the neutron star that forms in the interior of the collapsing star probably lead to sufficient energy deposits in the infalling matter to trigger an explosion. How exactly this occurs is a matter of current studies and debate. It appears to be the result of instabilities in infalling and expanding gas flows within the delicate balance of gravitational, nuclear reaction, neutrino interaction, and hydrodynamical flow energies.   Clumps and jets may be part of such explosions, as observed in several core-collapse supernova remnants. 

Nuclear reactions are expected to occur in the dense shock region approaching nuclear statistical equilibrium (NSE), hence producing Fe-group elements and also substantial amounts of radioactive $^{56}$Ni. At the same time, in regions nearer to the neutron star, neutrino interactions with nuclei occur in the \emph{$\nu$-process} and mainly liberate nucleons from nuclei, thus stimulating proton and neutron capture reactions of remaining nuclei. Further out, decomposition of in falling matter will also provide abundant free nucleons and $\alpha$-particles, and thus add an \emph{$\alpha$-rich} flavor to nuclear burning. Once the explosion sets in, material will expand and cool, and nuclear burning will \emph{freeze out}. Yet, the explosion will drive a shock through the outer envelope beyond the core, before these parts could take notice of the inner collapse, and some \emph{explosive nucleosynthesis} will occur in those shock-heated regions.  This explosive nucleosynthesis is characterized by short nuclear burning times which lead to large deviations from equilibrium and hydrostatic nuclear burning patterns.  
Altogether, one expects that abundant oxygen, silicon, and iron-group nuclei will be ejected, mixed with small but important contributions of heavy element products from the neutrino-driven wind zone.

Radioactive ejecta which could be observed with gamma-ray spectrometers include $^{56}$Ni and $^{44}$Ti, at estimated amounts of \about~0.1 and \about~few~10$^{-4}$~\Msol, respectively. Ejection of $^{56}$Ni is certified by the supernova light resulting from its decay energy, but amounts vary by almost three orders of magnitude \cite{2011Ap&SS.336..129N}). Again, proximity is key to gamma-ray observations, and more constraining due to the much smaller amounts of $^{56}$Ni compared to SNIa. $^{44}$Ti ejection is much less clear \cite{2006A&A...450.1037T}, as it is produced predominantly in the supernova's interior region very close to the mass cut separating ejecta from material which will end up in the central neutron star. 
One-dimensional models of Type II or Type~Ib supernovae obtain typical $^{44}$Ti yields of 5~10$^{-5}$~\Msol, with large (factors 2--4) variations with progenitor mass \cite{1996ApJ...464..332T}. Analyzing the impacts of deviations from spherical symmetry and the entropy-dependency of $\alpha$-rich freeze-out nucleosynthesis, an overall enhancement of $^{44}$Ti relative to $^{56}$Ni of a factor of \about~5 was considered plausible  \cite{1997ApJ...486.1026N}; but it remains to be shown that this holds for a realistic and detailed supernova model. Three-dimensional simulations of a nucleosynthesis network under a variety of plausible supernova-interior conditions \cite{2011ApJ...741...78M} obtain mass fraction variations over 6--8 orders of magnitude for $^{44}$Ti production, and thus reinforce that the specific supernova explosion trajectory in the $Y_e, \rho, T$ phase space  will determine the ejected amount of $^{44}$Ti. 
Measurements of $^{44}$Ti gamma-rays from all accessible young supernova remnants can clarify nucleosynthesis and thus the physical conditions in these inner supernova regions. 

 $^{44}$Ti decay results in gamma-ray emission in three lines: The first stage of its decay chain (half-life of 59~y ($\pm$0.3~y)) occurs predominantly (99.3\%) through capture of an electron into $^{44}$Sc, which de-excites by emission of the excitation-level energy through two X-rays at 68 and 78~keV. $^{44}$Sc itself $\beta$-decays quickly (with half life 3.7~h) to excited $^{44}$Ca, which in almost all cases de-excites through emission of the characteristic 1157~keV $\gamma$-ray line. 
The two low-energy lines are within the range of high-energy X-ray telescopes such as INTEGRAL's coded-mask telescopes, and the NuSTAR X-ray mirror telescope, while the gamma-ray line requires gamma-ray instruments such as COMPTEL aboard CGRO and SPI on INTEGRAL.

\begin{figure}
\includegraphics[width=0.53\textwidth]{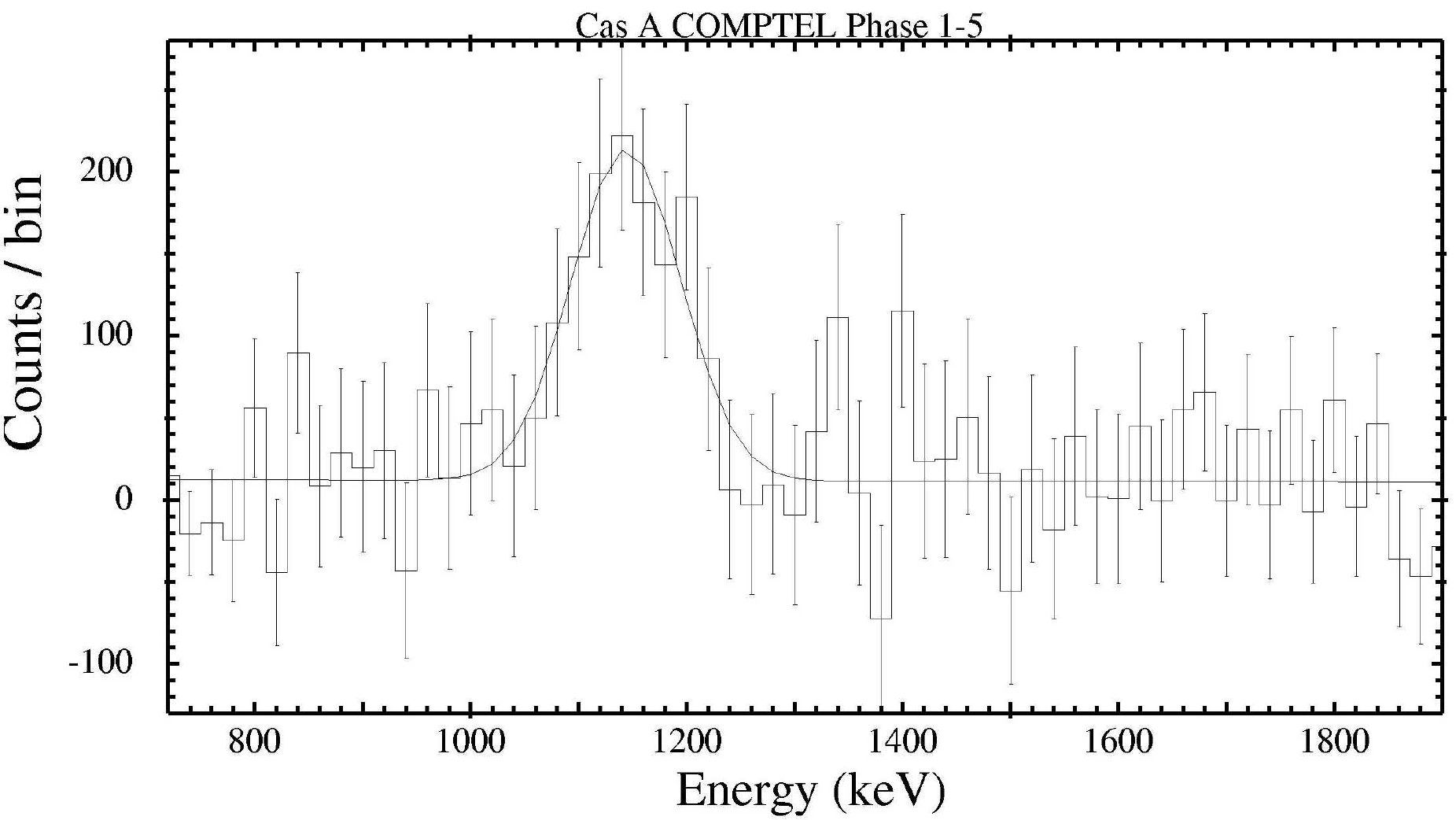}\includegraphics[width=0.45\textwidth]{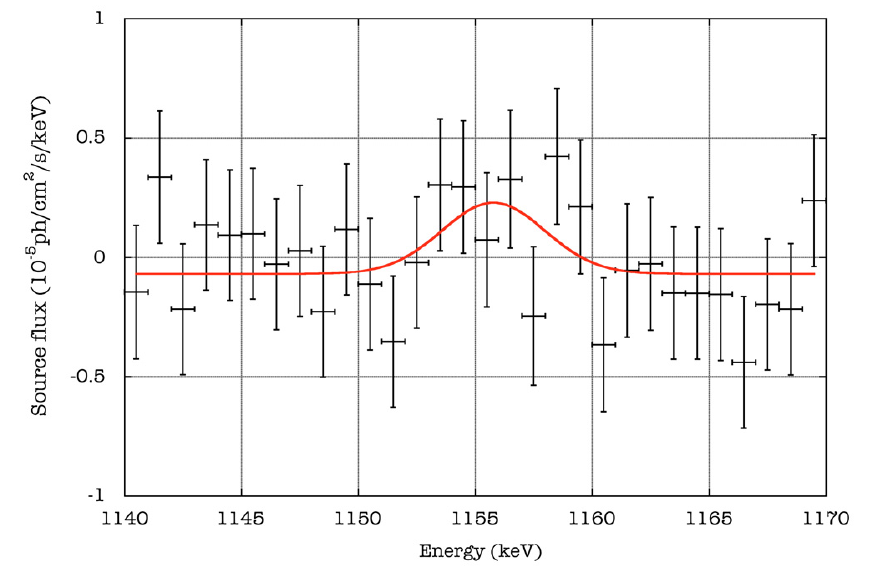}
\caption{The 1157 keV line from $^{44}$Ti decay. \emph{(Left:)} The $^{44}$Ti decay line at 1157 keV was discovered with COMPTEL. The detection was based mainly on imaging, detecting Cas A as a point source in the energy band of this line. \emph{(Right:)} INTEGRAL/SPI did not clearly detect the line; the continuous line represents a narrow line (instrumental width) with a flux of 1.6~10$^{-5}$ph~cm$^{-2}$s$^{-1}$, within uncertainty ($\pm$1.2) consistent with the consolidated flux of 2.5~10$^{-5}$ph~cm$^{-2}$s$^{-1}$. If we assume that the data do not show any line at the expected energy of 1157 keV, the Cas A line must be broadened by at least 500 km~s$^{-1}$ for statistical consistency (2$\sigma$) \cite{2009A&A...502..131M}.}
\label{fig:CasA-spec}
\end{figure}

The  Cas A supernova has become a key test case for core-collapse models. The remnant is at a distance of 3.4~kpc (+0.3/-0.1~kpc) \cite{1995ApJ...440..706R}, and its 5' diameter is suitable for observations also with modern telescopes throughout the electromagnetic spectrum, considering their often small fields of view. Its explosion is dated to year 1671 with a precision of \about~few years (see \cite{2004NewAR..48...61V,1997NuPhA.621...83H}). At present, the outer shock of the explosion blast wave is clearly seen in radio and X-ray bands, while the reverse shock that is expected cannot be clearly identified, but supposedly now has propagated through \about 2/3 of the remnant already, re-heating the remnant gas to X-ray emission temperatures. Although the age of Cas A is well beyond the 89~y decay lifetime of $^{44}$Ti, $^{44}$Ti radioactivity observations have become an important complement in the study of the Cas A supernova explosion. 

COMPTEL observations obtained discovery of $^{44}$Ti  from Cas A \cite{1994A&A...284L...1I} (Fig.~\ref{fig:CasA-spec}), and stimulated several followup measurements. 
Both INTEGRAL main telescopes are sensitive to $^{44}$Ti gamma-rays: The IBIS telescope obtained a flux value of  (2.3$\pm$0.5)~10$^{-5}$ph~cm$^{-2}$s$^{-1}$ \cite{2006ApJ...647L..41R} from the 68 and 78~keV lines. From the SPI spectrometer, an initial report of a detection consistent with expectations \cite{2007ESASP.622..105M} was corrected later towards an upper flux limit, owed to large variations of background during observations \cite{2009A&A...502..131M}.  This has been translated into an amount of $^{44}$Ti at the time of explosion of 1.6~x~10$^{-4}$~\Msol, with an uncertainty of ($^{+0.6}$/$_{-0.3}$)~\Msol. 
Thus the  $^{44}$Ti ejection from the Cas A supernova was \about a factor of three above predictions from models.
The non-detection of the 1157~keV decay line by INTEGRAL's SPI instrument \cite{2009A&A...502..131M} (Fig.~\ref{fig:CasA-spec}) suggests that $^{44}$Ti-rich ejecta move at velocities above 500~km~s$^{-1}$ \cite{2009A&A...502..131M}. As Doppler broadening increases with photon energy, this makes the high-energy decay line much broader than SPI's instrumental resolution, hence the signal-to-background ratio for this line degrades and may escape detection, less so the 68 and 78~keV lines which have been measured consistently from Cas A.
The NuSTAR imaging telescope has recently mapped the $^{44}$Ti emission morphology for Cas A (Harrison, \emph{priv.comm.,} 2013). This provides insights into supernova asymmetries, and revealed a rather extended, though incomplete, symmetry rather than major clumping of inner ejecta through $^{44}$Ti, which have been seen from other ejecta traces in the outer parts of the Cas A supernova remnant \cite{Vink:2004}. 

Supernova SN1987A in the LMC galaxy was \emph{the} event which had major influence on our current understanding of core-collapse supernovae. The detection of $^{56}$Co decay lines with the Gamma-Ray Spectrometer scintillation detector instrument on the Solar Maximum Mission  \cite{Matz:1988} was a surprise, as the supernova envelope had been expected to become transparent to inner nucleosynthesis products much later. This suggested major deviations from sphericity of the supernova explosion.  
INTEGRAL  observations in 2010/2011 provided a detection of SN1987A in just an energy band characteristic for the 68 and 78~keV lines from $^{44}$Ti decay \cite{GrebenevSN87A}. As with Cas A, the reported flux is considerably above expectations of $^{44}$Ti synthesis from models \cite{2011A&A...530A..45J,2002NewAR..46..487F,2010A&A...517A..51K}, again pointing to  non-spherical explosion models.

Considering the rate of core collapse supernovae in our Galaxy of \about~2/century \cite{2006Natur.439...45D}, one would expect to detect $^{44}$Ti emission from several objects along the plane of the Galaxy, if $^{44}$Ti ejection were a typical property of core-collapse supernovae. 
Both the COMPTEL and INTEGRAL suverys fail to detect the expected numbers of $^{44}$Ti sources. In spite of the low number of events involved, this is a significant observation  \cite{2006A&A...450.1037T,2006NewAR..50..540R}. Thus, also from the rate of  $^{44}$Ti emitting core-collapse supernovae, it seems that object to object variations must be large, and $^{44}$Ti ejection non-typical, probably much larger than current models predict in case they do happen. 
In non-spherical explosions with clumps and jets, $^{44}$Ti ejection could vary by more than an order of magnitude, as  explorations of the parameter space in density and temperature and their impact on nuclear reaction networks have shown \cite{2010ApJS..191...66M,2011ApJ...741...78M}.

\subsection{Radioactive Isotopes in the Interstellar Medium}\label{sources_ISM}

Radioactive nuclei ejected from cosmic sources of nucleosynthesis into interstellar space will accumulate and produce a diffuse glow, when their decay time is long compared to the time between ejection events. This is the case for $^{26}$Al and $^{60}$Fe from massive stars and their supernovae, given the radioactive decay times of 0.7 and 2.6 My, respectively, and massive-star evolution times of My with a core-collapse supernovae rate of \about~2 per century. Unlike in the case of the supernovae discussed above, the measured gamma-rays therefore cannot be attributed to a single object, and rather are due to the nucleosynthesis of a population or group of stars (see \cite{2009A&A...504..531V}). 

\begin{figure}
\centering
\includegraphics[width=0.68\textwidth]{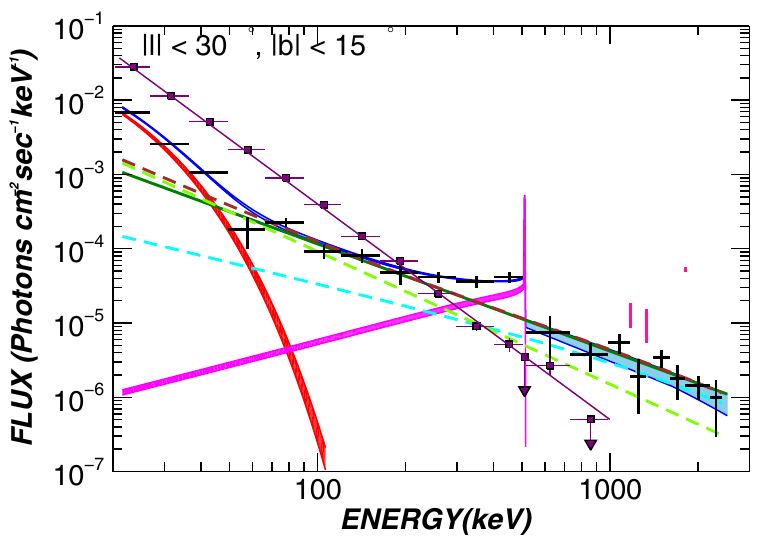}
\caption{INTEGRAL data integrated along the Galactic plane (longitude $-30<l<30^{\circ}$), showing continuum emission and the additional line feature from positron annihilation, as well as the lines at 511~keV, and from \Al, and $^{60}$Fe radioactive decay (adapted from \cite{2011ApJ...739...29B}). The SPI data points are shown as crosses, IBIS results  are shown with square symbols, and other lines (solid, dashed) show different components as expected to contribute to Galactic continuum emission (for details discussing the continuum components of Bremstrahlung, inverse-Compton, and superimposed stellar sources see \cite{2011ApJ...739...29B} and \cite{Strong:2007qe}). }
\label{fig:galridge_spec}
\end{figure}

\begin{figure}
\centering
\includegraphics[width=0.48\textwidth]{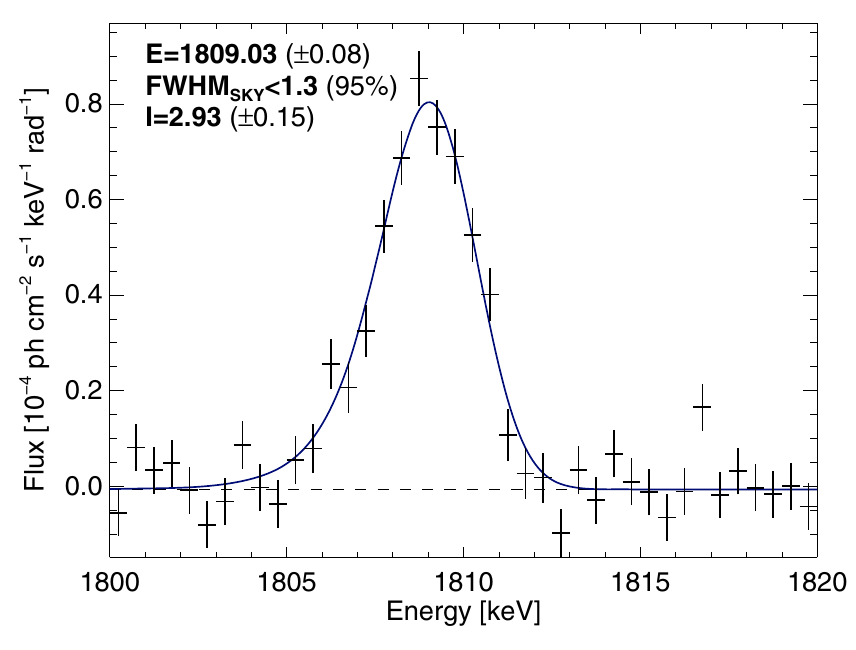}
\includegraphics[width=0.48\textwidth]{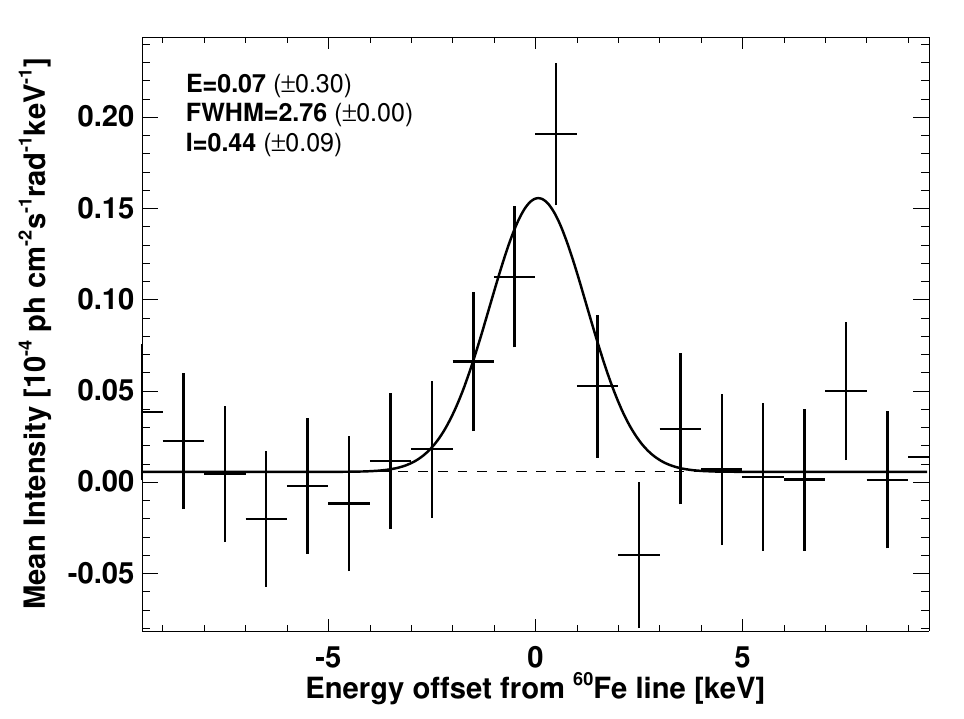}
\caption{The spectra measured by SPI from the entire plane of the Galaxy for $^{26}$Al \emph{(left)} \cite{2006A&A...449.1025D}, and $^{60}$Fe \emph{(right)} \cite{2007A&A...469.1005W}.}
\label{fig:Al-Fe_spectraSPI}
\end{figure}

The diffuse and extended gamma-ray emission of the Galaxy is dominated by a bright, power-law-type continuum, which is due to interactions of cosmic ray electrons with interstellar radiation and matter  (figure \ref{fig:galridge_spec},  \cite{2011ApJ...739...29B}. Superimposed are also seen a few line features, from positron annihilation (together with its lower-end continuum) at 511 keV, and from \Al at 1808.63~keV and $^{60}$Fe at 1773 and 1332~keV. 
These line emissions appear extended and diffuse in the Galaxy; this has been established for the bright \Al emission with sufficient signal for imaging, and inferred indirectly for $^{60}$Fe; the diffuse glow of positron annihilation emission, which should go hand in hand with nucleosynthesis of isotopes on the proton-rich side of the stability valley of isotopes, such as $^{26}$Al or $^{44}$Ti or $^{56}$Ni, has not yet been detected unambiguously from the Galaxy as a whole, and also not from the prominent $^{26}$Al-bright massive-star regions, but rather is found concentrated in a bright and extended emission region associated with the Galaxy's bulge region.
It is these lines which will be discussed in more detail in the following.

\subsubsection*{$^{26}$Al in the Galaxy}
The \Al line from the interstellar medium in the Galaxy was the first nuclear line detected from outside the solar system \cite{Mahoney:1984}. It had been attributed to current nucleosynthesis in our Galaxy, although the source remained unclear for a while between novae, Wolf-Rayet stars, supernovae, and even solar system origins. The COMPTEL gamma-ray survey then mapped \Al emission across the sky \cite{1995A&A...298..445D,1999A&A...344...68K}, and found rather clumpy emission extended along the entire plane of the Galaxy. This was interpreted as massive stars being the dominant sources, evolving in groups and leading to local emission peaks when stars reach Wolf-Rayet phases and first core-collapse events, while novae were considered less important because their more-frequent ejection events should produce a smoother and more centralized glow of the Galactic plane \cite{1996PhR...267....1P}. 
 
\begin{figure}
\centering
\includegraphics[width=1.0\textwidth]{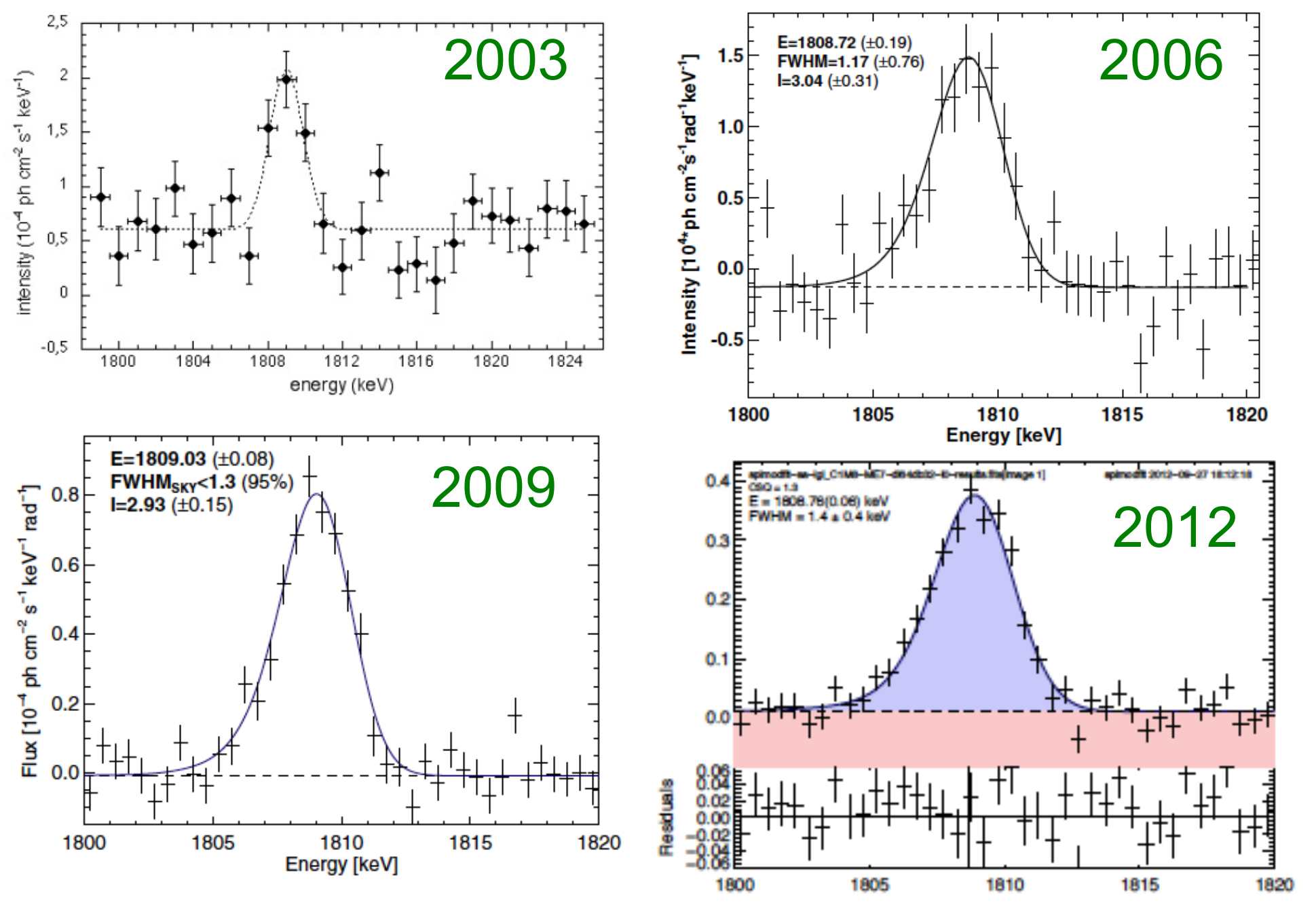}
\caption{The \Al line measurements with SPI during the INTEGRAL mission.}
\label{fig_Al-spectra_years}
\end{figure}

SPI confirmed the interstellar glow throughout the Galaxy of $^{26}$Al gamma-rays early in the INTEGRAL mission, and has been refining those measurements since. 
Figure~\ref{fig_Al-spectra_years} shows the  \Al signal measured with INTEGRAL/SPI from the entire Galaxy, adopting spatial distribution on the sky according to the skymap based on the COMPTEL measurements \cite{Pluschke:2001c}. Shown are results from different datasets over the years, the latest result using also multiple-detector events and data up to orbit number 1142 (February 2012). The \Al line parameters of intensity, line position, and width, are all consistent with earlier results based on single-detector events only. The statistical precision increases from the larger number of events, although some additional uncertainty from background modeling arises. Nevertheless, the \Al emission is seen at 32~$\sigma$ significance.
This  signal from  $^{26}$Al has been studied in detail now from the Galaxy at large \cite{2006A&A...449.1025D,2009A&A...496..713W}, see Fig.~\ref{fig:Al-Fe_spectraSPI}, and in specific, localized regions (see below).  
In the inner Galactic ridge, a flux at 1808.63~keV of (2.63~$\pm$0.2)~10$^{-4}$ph~cm$^{-2}$s$^{-1}$rad$^{-1}$ has been derived \cite{2010A&A...522A..51D}. 

The total mass of $^{26}$Al in the Galaxy had been inferred from earlier COMPTEL measurements to be 2--3~\Msol \cite{1996PhR...267....1P}. In such a mass determination, one must adopt a spatial source distribution to resolve the distance uncertainty, when converting a measured gamma-ray flux into a quantity of isotopes present in the Galaxy. Generally, large-scale models of Galactic structure or sources have been used, such as exponential-profile disk, azimuthally-symmetric nested rings, or spiral-arm models inferred from other observations. Yet, the irregularity of the large-scale emission as seen by COMPTEL from Galactic $^{26}$Al already suggested that the massive-star population in the Galaxy may be more clumpy than such large-scale models describe. The localized $^{26}$Al emission seen from the Cygnus and Sco-Cen regions \cite{2009A&A...506..703M,2010A&A...522A..51D} supports this view. Accounting for such localized enhancements, with more-advanced INTEGRAL measurements the Galactic mass of $^{26}$Al has been re-determined as ranging from 1.5 to 3.6~\Msol, considering all uncertainties.

Using this Galaxy-wide amount of $^{26}$Al together with $^{26}$Al yields for massive stars across the entire mass range, and an initial-mass distribution, one can derive the total population of stars which corresponds to measured $^{26}$Al gamma-rays \cite{2006Natur.439...45D}. As massive stars above \about~8-10~\Msol all are believed to end their evolution as core-collapse supernovae, this corresponds to a determination of the \emph{Galactic} rate of core-collapse supernovae. We obtain a value of 1.54 ($\pm$0.89) supernovae per century, or a supernova every 65 years (somewhat lower than the value published in \cite{2006Natur.439...45D} from $^{26}$Al, due to accounting for foreground emission now attributed to the nearby Scorpius-Centaurus sources, rather than distant Galactic $^{26}$Al.).  
The importance of this measurement of the Galactic supernova rate derives from the underlying method. Using penetrating gamma-rays from radioactivities ejected by supernova related sources throughout the current Galaxy, this approach does not suffer from corrections for occulted sources, such as e.g. a supernova rate determination based on O and B star counts does. Other measurements inherently assume negligible possible differences between our Galaxy and other galaxies, as they measure core-collapse supernova related objects in Milky-Way-like galaxies which are seen face-on and hence free from occultation-bias of distant parts of our own Galaxy. It is interesting that the $^{26}$Al-based supernova rate determination agrees with measurements undertaken with those alternative approaches, and falls into the lower-value part among all those measurements (as discussed in \cite{2006Natur.439...45D}).

\subsubsection*{The $^{26}$Al Line Shape}
\begin{SCfigure}
\centering
\includegraphics[width=0.6\textwidth]{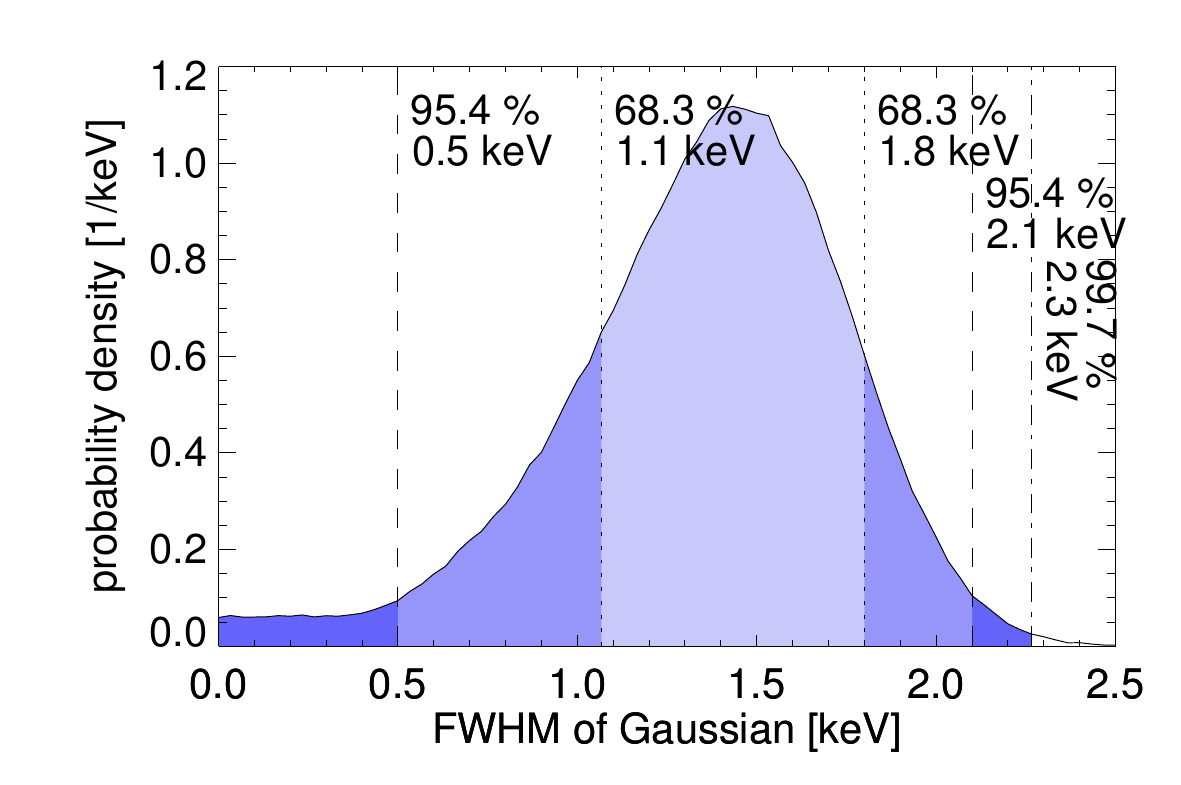}
\caption{The astrophysical  \Al line width constraint. (see text for a discussion).}
\label{fig_Al-width_2012}
\end{SCfigure}

With fine Ge-detector spectroscopy, the shape of the \Al is resolved, and may thus be studied in terms of astrophysical signatures.
Figure~\ref{fig_Al-width_2012} shows the probability distribution of the line width for an additional Gaussian line broadening beyond the instrumental resolution alone. An astrophysical line broadening was discovered, with a value of 1.4~keV ($\pm$0.3~keV), which corresponds to 175~km~s$^{-1}$ ($\pm$45~km~s$^{-1}$) in velocity space. For comparison, the expectations from averaging over different velocities from large-scale rotation of the Galaxy are about 1.35~keV, and compatible with this measurement.

\begin{figure}
\centering
\includegraphics[width=0.9\textwidth]{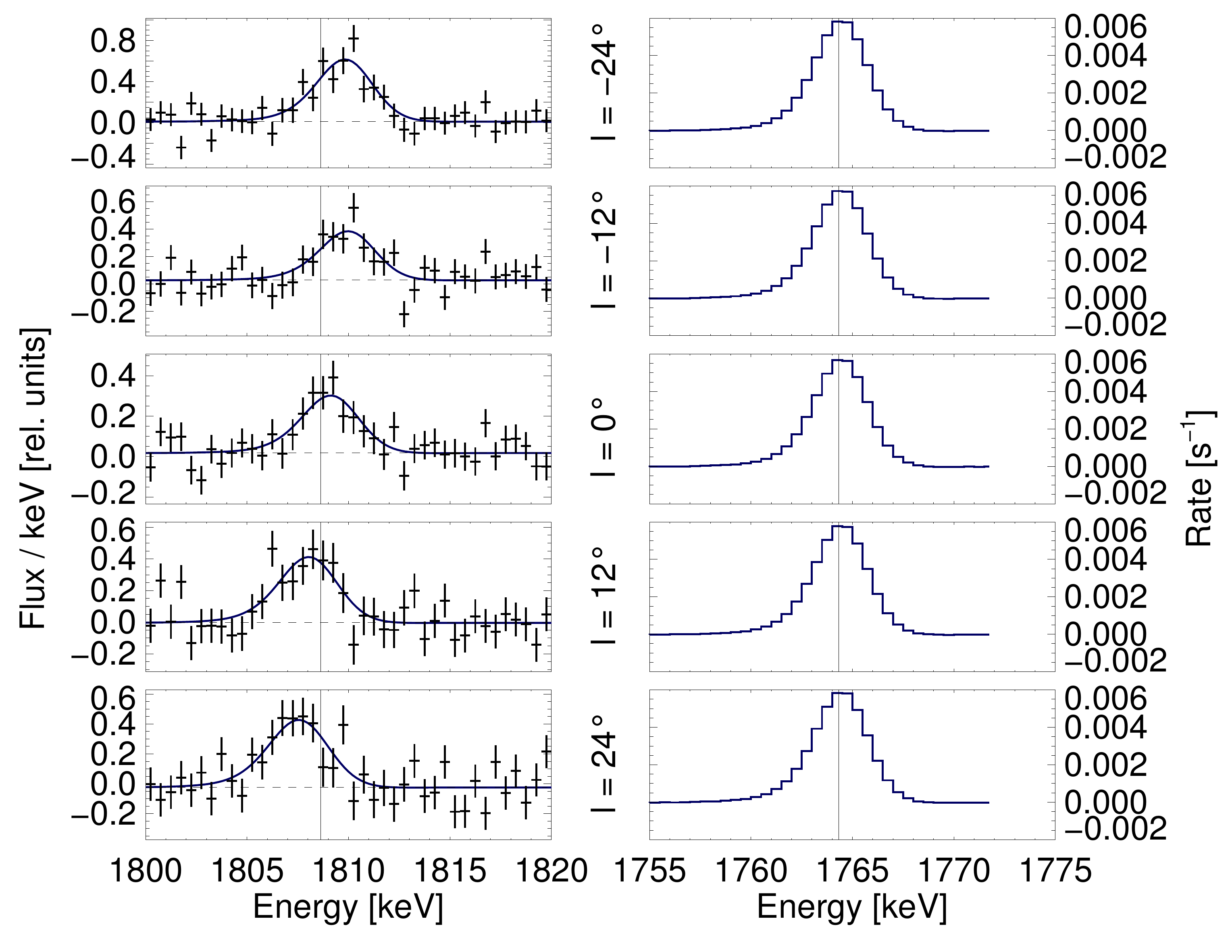}
\caption{The \Al line measurements with SPI along the plane or the Galaxy, for different lines of sight. The systematic Doppler shift from large-scale rotation in the galaxy can be seen clearly (\emph{left}). A nearby instrumental line at 1764 keV (\emph{right}) does not show differences in line position, from the same data.}
\label{fig_Al-spectra_plane}
\end{figure}

The large-scale Galactic rotation shows up as  a line centroid shift in the measurements of \Al: Bulk motion differs for different lines of sight along the plane of the Galaxy, with systematic variations leading to a blue shift of the \Al line towards the 4$^{th}$ quadrant of the Galaxy, and a red shift towards the 1$^{st}$ quadrant, correspondingly. The Doppler shift signature was indicated in early results, and is now consolidated (see Fig.\ref{fig_Al-spectra_plane}). 
This allows to study motion of hot interstellar gas as it is shaped around massive-star regions, and compare its dynamics to dynamics of stars and gas as we know it otherwise. For example, CO line emission has been used to trace the molecular gas in the Galaxy \cite{2001ApJ...547..792D}. Here, the Galactic ridge was recognized clearly, with peculiar and high cloud dynamics in the vicinity of the Galaxy's central supermassive black hole. As star formation occurs from dense clumps in molecular clouds, \Al ejected from recently-formed stars could show how closely its kinematic motion can be related to the current molecular cloud population.  

\begin{figure}
\centering
\includegraphics[width=0.78\textwidth]{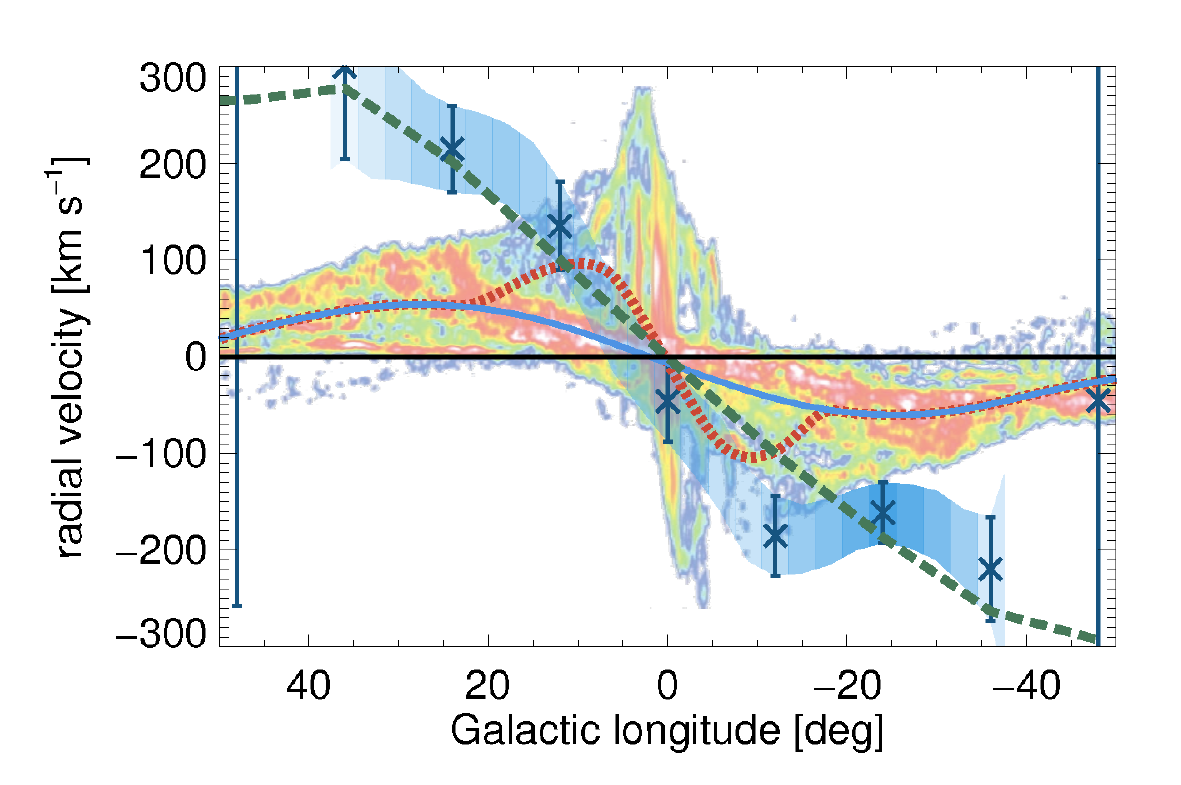}
\caption{The  $^{26}$Al line  Doppler-shift signature along the inner Galaxy, as plausible from large-scale Galactic rotation, and compared to CO data \cite{Dame:2001} (color). The data points are derived from independent longitude segments and show the velocity of $^{26}$Al-enriched interstellar gas; the blue-hatched area shows the range of uncertainty in these values from imaging resolution. The solid blue line shows expectations from \Al sources moving with molecular clouds as seen in CO data, the red-dashed line shows how locations of \Al sources in the Galaxy's bar would affect the kinematic signature.  The dashed line matching the \Al  data points shows expectations from a model based on \Al being ejected from the leading edges of inner Galactic spiral arms, with a net bulk velocity of \about~200~km~s$^{-1}$ due to asymmetries of the superbubbles created by massive star groups; see Kretschmer et al., 2013, submitted, for more detail.}
\label{fig:26Al_long-vel}
\end{figure}

The longitude-velocity diagram as  derived from INTEGRAL \Al measurements is shown in Fig.~\ref{fig:26Al_long-vel}. The derived velocities extend from \about~-200 to \about~+200~km~s$^{-1}$ at the extremes (tangential directions to Galactic-ridge directions with largest relative motion with respect to the Sun, at longitudes $\pm$30\degree. 

It is apparent that the \Al velocities extend over higher velocities on large scale than known for other sources and in particular for cold, molecular gas (also shown in Fig.~\ref{fig:26Al_long-vel} in color). This appears surprising. If we consider \Al sources across the Galaxy traced by free electrons produced by ionizing starlight, which has been mapped through pulsar dispersion measurements, we would expect a longitude-velocity trend as shown by the thick (blue) line in Fig.~\ref{fig:26Al_long-vel}. But this representation of free electrons according to \cite{2002astro.ph..7156C} is deficient in the inner Galaxy, from an observational bias of available pulsar data. Thus it may not be surprising that this apparently does not represent \Al observations of motions, providing a hint that some inner-Galaxy sources may be missing in this free-electron model. Adding an additional \Al source population also within a Galactocentric radius of 2~kpc (where the free-electron map is \about~empty), and applying a density structure according to the Galaxy's inner bar \cite{2005ApJ...630L.149B} plus rotational behavior as extrapolated towards the inner Galaxy from available measurements \cite{2009PASJ...61..227S}, the \Al kinematic data are more closely reproduced (Kretschmer et al., submitted).  The generally-larger velocities seen in \Al sources suggest that preference towards the direction of Galactic rotation may be given to massive-star ejecta as they leave their sources. It remains to be shown if more detailed modeling of star formation along the Galaxy's bar and inner spiral arms  can provide an explanation of the observed \Al velocities.

Surprisingly, \Al enriched gas appears to show systematically larger velocities than expected. This suggests that the ejection of nucleosynthesis material from massive-star sources occurs into surrounding interstellar medium which has been shaped by massive-star winds and the gas accumulation in spiral arms in peculiar ways (Kretschmer et al. 2013, in press). 


\subsubsection*{$^{26}$Al and $^{60}$Fe from Massive Stars}

Long searched for,  gamma-rays from decay of $^{60}$Fe  were discovered from the plane of the Galaxy in 2005 by RHESSI \cite{Smith:2004} and INTEGRAL \cite{2007A&A...469.1005W}, see Fig.~\ref{fig:Al-Fe_spectraSPI}. 
This signal was found to be much fainter than the one from $^{26}$Al, although the same massive-star populations are the most-plausible sources for both those relatively long-lived isotopes which decay long after they have been injected into interstellar space from their sources. 

\begin{figure}
\centering
\includegraphics[width=0.68\textwidth]{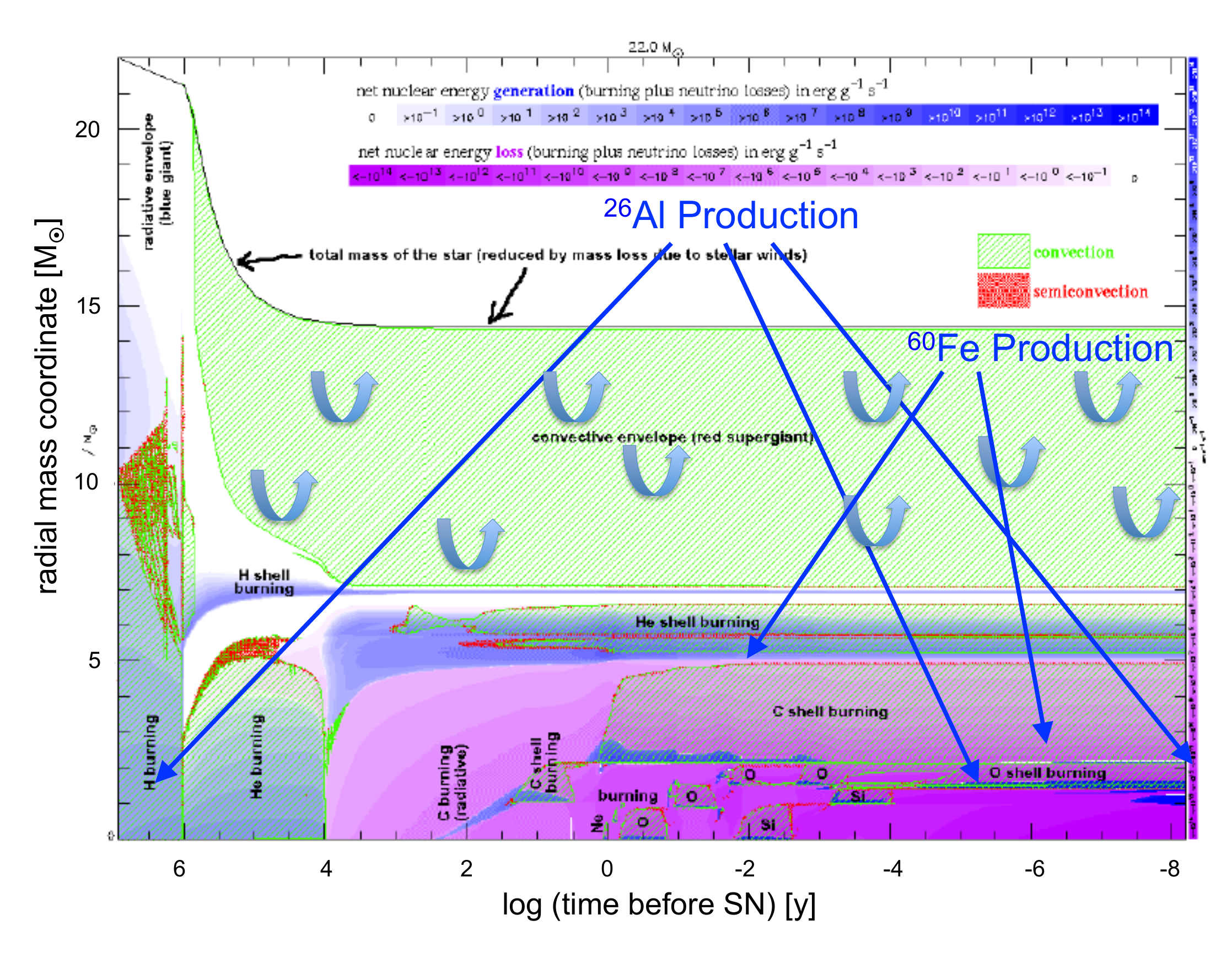}
\caption{The interior structure of massive stars is complex towards their later evolution: Intermittent shell burning occurs, simultaneously and in addition to core nuclear burning. This \emph{Kippenhahn diagram} shows the evolution of a massive-star interior towards the supernova in a logarithmic time scale (adapted from \cite{2003NuPhA.718..159H}, for a star of initial mass 22~\Msol). The production sites of $^{26}$Al and $^{60}$Fe are indicated, as well as the mass loss from stellar wind in the Wolf-Rayet phase. }
\label{fig:Al-Fe_nucleosynthesis}
\end{figure}

The \Al yields of massive stars used in above considerations are based on models of stellar evolution and supernovae (see \cite{Diehl:2006d} for a summary and references).    
Fig.~\ref{fig:Al-Fe_nucleosynthesis} shows  the complex interior structure of massive stars, as they evolve from initial core hydrogen burning towards the final core-collapse supernova. Nuclear-burning and convective regions are indicated (hatched).  Within the star, at later phases, intermittent shell burning occurs simultaneously and in addition to core nuclear burning of the different fuels, He, C, O and Si burning.  The locations of $^{26}$Al (and also $^{60}$Fe) synthesis are indicated (arrows): $^{26}$Al is produced in the hydrogen-burning phases from initial $^{25}$Mg, and later in the carbon and neon burning core and shell as proton-release reactions allow further processing of remaining $^{25}$Mg, plus during the supernova explosion as explosive Ne/C burning. Only $^{26}$Al from core hydrogen burning is mixed into the envelope and thus ejected during the intense stellar-wind phase. $^{60}$Fe is expected to be synthesized from successive neutron capture reactions on $^{54}$Fe, as neutron release reactions such as $^{13}$C($\alpha$,n) and $^{22}$Ne($\alpha$,n) are activated; but here, a subtle balance of neutron capture and convective transport away from neutron-rich regions is required to ensure that $^{60}$Fe is not destroyed by further neutron captures. Reaction rates for these neutron captures on unstable and n-rich Fe isotopes are all being re-assessed by current and planned laboratory experiments (see \cite{Wiescher:2012nx} for a discussion). All $^{26}$Al and $^{60}$Fe produced within the star after the main sequence phase is released only with the supernova explosion, together with additional contributions from explosive burning in the supernova itself. Note that stellar evolution uncertainties also arise from nuclear reaction rate uncertainties of the 3$\alpha$ and $^{12}$C($\alpha,\gamma)^{16}$O reactions, and in particular affect the yields of $^{60}$Fe \cite{Tur:2010}. 

If we consider the likely formation of a \emph{group} of massive stars from a dense molecular-cloud core, their concerted release of $^{26}$Al and $^{60}$Fe radioactive isotopes occurs over a period of \about~3 to 20~My after star formation \cite{2009A&A...504..531V}. The measurement of $^{26}$Al and $^{60}$Fe radioactivity gamma-rays, therefore, is a tool to verify our \emph{stellar-mass averaged} predicted yields from models of massive-star evolution and nucleosynthesis. In particular, since the same massive stars are plausible producers of both those isotopes, yet from different regions and epochs inside the stars, the measurement of the ratio of $^{26}$Al to $^{60}$Fe emission provides a valuable tool, as systematic uncertainties in the stellar populations themselves (richness, distance) cancel in such ratio.  

The $^{60}$Fe gamma-ray line emission occurs in a cascade of two gamma-ray lines, at 1172.9 and 1332.5~keV, with approximately-equal intensities (99.85\% of decays produce the 1172.9~keV line, versus 99.98\% for the 1332.5~keV line; a 59~keV line is emitted in 2\% of decays from a transition in $^{60}$Co). In none of these lines, separately analyzed in SPI single-detector and multiple-detector events, a clearly-convincing celestial gamma-ray line signal could be seen. Combining line intensities in both these lines, the spectrum shown in Fig.~\ref{fig:Al-Fe_spectraSPI} has been derived, which shows $^{60}$Fe emission from the sky with a combined significance of \about~5$\sigma$. In any case, the total gamma-ray brightness in $^{60}$Fe decay is substantially below the one seen from $^{26}$Al. The measured gamma-ray brightness ratio is \about~15\%, with an uncertainty of \about~5\%. For $^{60}$Fe, no imaging decomposition could be derived yet due to this low brightness. 
In steady state, this brightness ratio thus constrains massive-star interiors globally for an average over all massive-star groups throughout the Galaxy. For stellar groups at specific ages, the steady-state assumption does not apply, however, and in particular the re-assessed decay time of $^{60}$Fe of 3.8~My  \cite{Rugel:2009} (the formerly-used value was 2.2~My) must be taken into account for determinations of the isotope ratio (see discussion of \emph{specific regions} below).

From nucleosynthesis models of massive stars and supernovae, it appears that more $^{60}$Fe would be expected, although currently-predicted yields are still in agreement with the observations, within quoted uncertainties. Note that $^{60}$Fe nucleosynthesis may also occur in a rare subtype of SNIa explosions at high yields of \Msol  \cite{Woosley:1994}, although a single bright source being responsible for observed $^{60}$Fe gamma-rays appears unlikely (the RHESSI and SPI detections were derived assuming extended emissions along the Galactic plane). $^{60}$Fe yields are particularly high (as compared to $^{26}$Al yields) in massive-star models for the upper mass ranges above \about~60~\Msol \cite{Limongi:2006a,Limongi:2006qf}. So, even considering the uncertainties of the gamma-ray measurement, the $^{60}$Fe/$^{26}$Al ratio is lower than expected. 
Possible causes could be suppression of $^{60}$Fe by more-efficient neutron capture reactions, leading to neutron-richer Fe isotopes. Alternatively, $\beta$-decay of $^{59}$Fe could occur at a higher rate than inferred from current nuclear theory. Both reaction types are being investigated in laboratory experiments at the present time. Note that the $\beta$-decay lifetime of the $^{60}$Fe isotope had been re-determined in 2008 \cite{2009PhRvL.103g2502R}, and found to be substantially longer, with (exponential) lifetime increased from 2.15 to 3.8~My (T$_{1/2}$ from 1.5 to 2.6~My).
Another cause of lower Galaxy-wide $^{60}$Fe glow could be that the explosion as a supernova does not occur throughout the entire range of stellar masses, but deviations from the standard initial-mass distribution on the high-mass end of massive stars $\geq$60~\Msol, or 'islands of explodability' in that mass range occur. 
The observational bound of \about~15\% for the $^{60}$Fe/$^{26}$Al gamma-ray intensity ratio provides a significant constraint to be met by massive-star models as a whole, including aspects of stellar structure, nucleosynthesis, and evolution towards supernovae.

Particularly interesting would be the $^{60}$Fe/$^{26}$Al ratio for source populations of specific ages, as the ratio varies significantly due to wind-released $^{26}$Al before any core-collapse supernova would eject $^{60}$Fe and more $^{26}$Al \cite{2009A&A...504..531V}. $^{60}$Fe is exclusively released in supernovae, although predominantly produced in the late shell-burning phase before the collapse of the core. Only the Cygnus region appears within INTEGRAL's sensitivity range for this, however \cite{2010A&A...511A..86M}. 

\subsubsection*{Special Regions}
\begin{figure}
\centering
\includegraphics[width=0.98\textwidth]{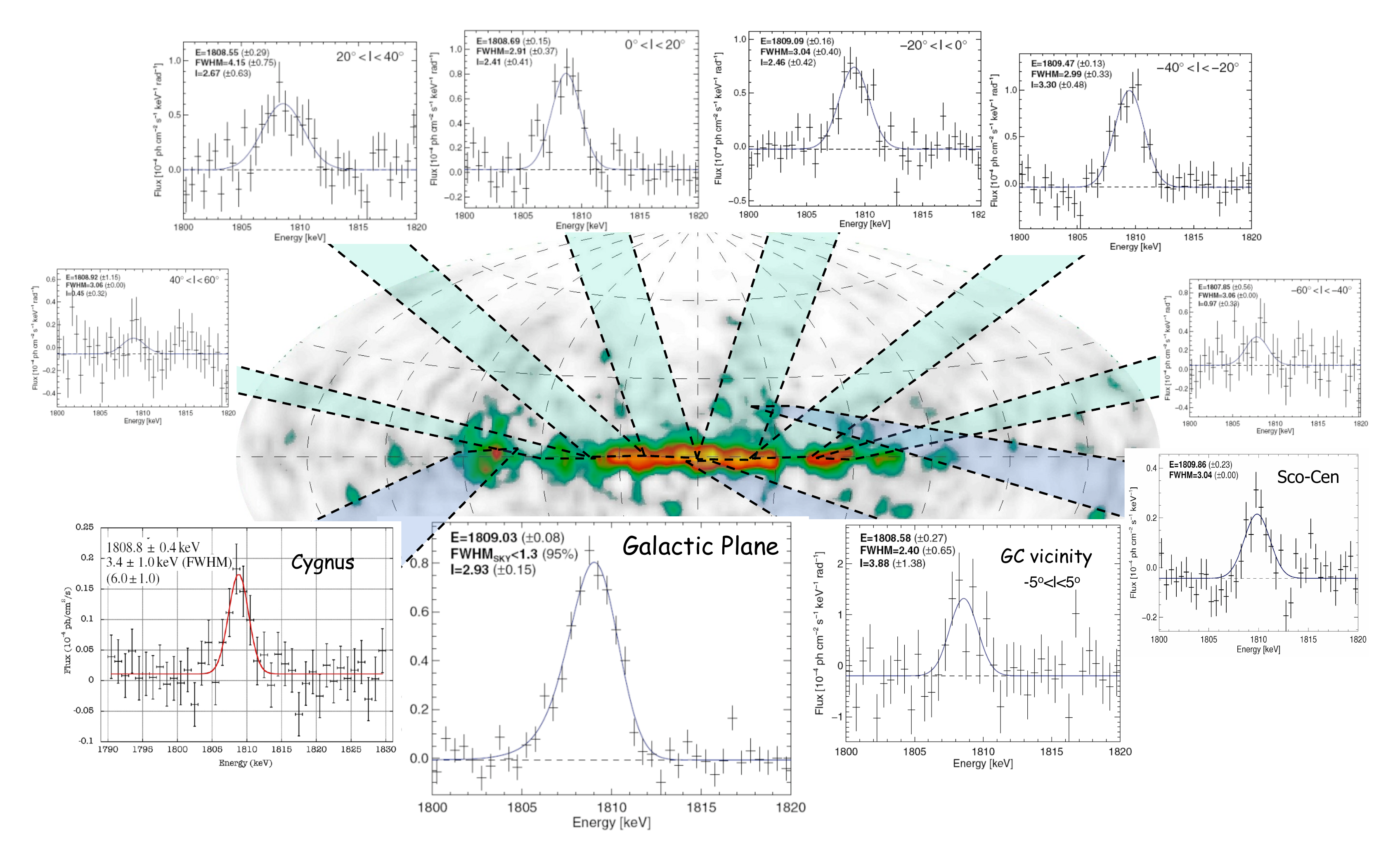}
\caption{The spectra showing the \Al gamma-ray line (E$_{lab}$~1808.63~keV) as measured by SPI from different regions along the plane of the Galaxy for $^{26}$Al \cite{2009A&A...496..713W}. }
\label{fig:Al_region_spectraSPI}
\end{figure}

\begin{SCfigure}  
\centering 
\includegraphics[width=0.65\textwidth]{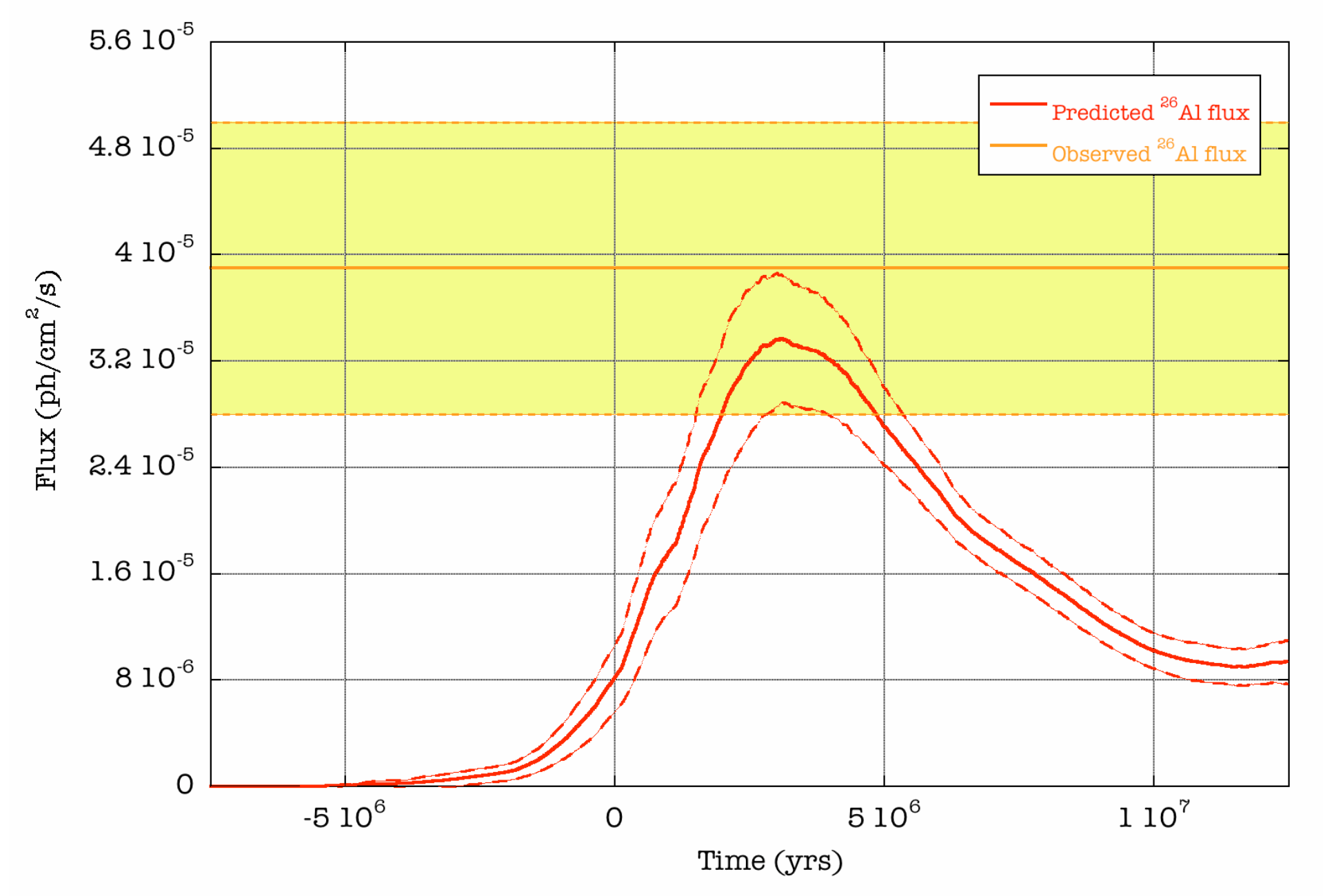}
\caption{The time history of \Al production in the Cygnus complex, as compared to the gamma-ray observations.  (see text).
}
\label{fig7:al_cygnus} 
\end{SCfigure}

$^{26}$Al is bright enough to also be seen from localized regions along the Galactic plane hosting many massive stars (Fig.~\ref{fig:Al_region_spectraSPI}). For 
a specific source region, distance uncertainties can often be resolved from other observations, and, moreover, the stellar population is determined through star catalogues. Therefore, a  comparison of predicted versus observed amounts of \Al can be made (see Fig.~\ref{fig7:al_cygnus}), and provide a more specific test for massive-star models than can be obtained from Galaxy-wide analysis as discussed above. With deeper exposure, this became possible in the late INTEGRAL mission for the Cygnus \cite{2010A&A...511A..86M}, Carina \cite{2012A&A...539A..66V}, and Scorpius-Centaurus \cite{2010A&A...522A..51D} regions;  for other candidate locations of massive-star groups INTEGRAL's sensitivity is insufficient, due to their fainter \Al emission.

The brightest individual \Al emission region appears to be the Cygnus region. Between 6 and 9 OB associations potentially contribute to the signal, whose ages range from 2.5 to 7.5~My \cite{2002A&A...390..945K}. Probably the Cyg OB2 association dominates by far  \cite{Martin:2010b}: About 120 stars in the high-mass range (20--120~\Msol) have been identified to relate to Cyg OB2; the other associations typically are ten times smaller. The age and distance of Cyg OB2 is 2.5~My and 1.57~kpc, respectively. Because of its young age,  stellar evolution even for the most-massive stars should still not be completed, and contributions from core-collapse supernovae to \Al production should be small or absent. Instead, Wolf-Rayet-wind ejected \Al from hydrostatic nucleosynthesis may be assumed to dominate, currently originating from Cyg~OB2 stars. 

The \Al emission towards the Cygnus direction has been measured with INTEGRAL in the longitude interval [70\degree,96\degree] as a total of $\sim$6~10$^{-5}$ph~cm$^{-2}$s$^{-1}$ \cite{2004ESASP.552...33K,2009A&A...506..703M}.  Accounting for a large-scale galactic background, the contribution from the \emph{Cygnus complex} is  $\sim$3.9~10$^{-5}$ph~cm$^{-2}$s$^{-1}$ \cite{2009A&A...506..703M}. Accumulation of the expected \Al production from the stellar census has always shown that \Al gamma-rays seemed 2--3 times brighter than predicted. 
In Fig.~\ref{fig7:al_cygnus}, the horizontally-shaded area presents the range given by the \Al gamma-ray data, the dashed lines bracket the uncertainty range of predictions from recent massive-star models through population synthesis. Expectations from such populations synthesis are on the low side of observed \Al gamma-rays, and predict the main \Al ejection still to come within the next \about~3~My.  
While much of this discrepancy could be assigned to the occultation of stars by molecular clouds \cite{2002A&A...390..945K}, some under-prediction ($\sim$25\% for solar metallicity) remained, though still within uncertainties of both observation data and models \cite{2010A&A...511A..86M}. The metallicity dependence of Wolf-Rayet-phase  \Al yields results for Cygnus region metallicity in a revised \Al prediction by a factor \about~4 below the measurement (see Fig.~\ref{fig7:al_cygnus} at time=0) \cite{2010A&A...511A..86M}.

For a young and active region of massive-star action, one may plausibly assume that the interstellar medium would be peculiar and probably more dynamic than in a large-scale average. With the fine spectroscopic resolution of the INTEGRAL measurements, therefore initial hints for a broadened \Al gamma-ray line were tantalizing. With better data, it turns out that the \Al line seen from the Cygnus region is compatible with the laboratory energy (i.e. no bulk motion exceeding tens of km~s$^{-1}$) and with instrumental line width (i.e. no excessive Doppler broadening beyond $\sim$200~km~s$^{-1}$ \cite{2009A&A...506..703M}. Note that \Al ejection from Wolf Rayet winds would be with $\sim$1500~km~s$^{-1}$, decelerating as circum-stellar gas is swept up.

The nearby region corresponding to stars from the Scorpius-Centaurus association has been discriminated against the Galactic emission in \Al \cite{Diehl:2010}. This stellar association and its subgroups are located at a distance of about 100--150~pc \cite{1992A&A...262..258D,1999AJ....117..354D,1999AJ....117.2381P}. Several subgroups of different ages (5, 16, and 17~My, with typical age uncertainties of 1--2 My; \cite{1989A&A...216...44D}, \cite{2008ApJ...688..377S}) have been identified. Similar to the Orion region, this promises that even modest spatial telescope resolution such as from SPI (2.7$^{\circ}$) could reveal \Al emission displaced from its sources, and thus teach us about ejecta flows. With SPI, \Al emission from the direction overlapping the current location of the Upper-Sco group of stars could be discriminated against the large-scale Galactic \Al emission, due to its favorable location about 20\degree above the plane of the Galaxy \cite{2010A&A...522A..51D}. 
The \Al ejected from the Scorpius-Centaurus stars should in fact be distributed over a large region on the sky. But the distribution might help to distinguish contributions from different-age subgroups, thus helping to study the scenario of triggered star formation. 

The \Al line from  the Scorpius-Centaurus region source may be slightly  blue-shifted:  a centroid energy of 1809.46~keV ($\pm$0.48 keV) has been found, and  implies a blue shift of $\sim$0.8~keV corresponding to bulk streaming towards the Sun at about (137$\pm$75)~km~s$^{-1}$. 
Interestingly, measurements of hot gas in the solar cavity also suggested gas inside the local cavity streaming from this general direction towards the Sun \cite{2009SSRv..146..235F}. 

The Orion region is the most-nearby region of massive stars, at a distance of $\sim$450~pc \cite{2008hsf1.book..459B,1989ARA&A..27...41G}. Its location towards the outer Galaxy and at Galactic latitudes around 20\degree is favorable, as potential confusion from other Galactic sources is negligible. The dominating group of massive stars is the Orion OB1 association \cite{1994A&A...289..101B} with three major subgroups of different ages, one oldest subgroup \emph{a} at 8--12~My, and two possibly coeval subgroups \emph{b} (5--8~My) and \emph{c} (2--6~My); subgroup \emph{d} is the smallest and youngest at 1~My or below. Subgroup c holds the most massive stars, about 45 in the mass range 4--120~\Msol. These groups are located on the near side of the Orion A and B molecular clouds, which extend from 320 to 500~pc distance away from the Sun, and span a region of $\sim$120~pc perpendicular to our viewing direction.  

 \Al emission from Orion was seen with COMPTEL (a gamma-ray flux of 7~10$^{-5}$ph~cm$^{-2}$s$^{-1}$ at  5$\sigma$ significance, \cite{2002NewAR..46..547D}), and appears quite extended and not concentrated near the Orion OB1 association. A huge interstellar cavity extends from the Orion molecular clouds towards the Sun, the 'Eridanus' cavity, which extends over almost 300~pc  and is seen in X-ray emission \cite{1993ApJ...406...97B}.   \Al ejected from current-generation stars would find a pre-shaped cavity directing the flow of ejecta into it, rather than towards the dense remains of the parental molecular clouds on the far side of the OB association.
With INTEGRAL's spectrometer the \Al line centroid (bulk motion towards the Sun?) and width (\Al fractions moving within the hot cavity versus \Al deposited at the cavity walls?) may provide interesting high-resolution spectroscopy constraints, from observations taken in 2012/2013.

\subsection{Positron Annihilation}\label{positrons}
Nuclei such as the above-discussed $^{26}$Al, $^{44}$Ti, and $^{56}$Ni isotopes decay through $\beta^+$-decays, thus releasing positrons into interstellar space, with typical energies of \about~MeV. Colliding with their antiparticles, the electrons, in the interstellar medium, annihilation photons are released, with a total energy equivalent to the rest mass of the two leptons of 1.022~MeV plus their kinetic energy. From momentum and spin conservations, the photon spectrum is dominated by photons at 511~keV, plus a continuum with a maximum energy of 511 keV from annihilation through the ortho-positronium atom made of the two leptons as an intermediate state. There are other candidate sources of interstellar positrons, such as magnetized and rotating neutron stars and accreting binary systems, which are expected to release electron-positron plasma with much higher energies (\about GeV), and dark-matter particle interactions may also plausibly contribute to produce positrons (for a review see \cite{Prantzos:2011}). 
The positron annihilation gamma-ray line was the first cosmic gamma-ray line ever detected, in 1972, with a low-resolution NaI detector instrument \cite{1972ApJ...172L...1J}, and later identified as annihilation line through the Ge detector measurement providing sufficient spectroscopic precision for line identificantion\cite{1978ApJ...225L..11L}. 

Apparently time-variable annihilation emission was pursued for a while, hoping for a key diagnostic of pair plasma in compact sources \cite{1982ApJ...260L...1L,1986ApJ...302..459L}. But most of the apparent variability was later found to appear because different instruments with different field-of-view sizes recorded different fractions of the diffuse galactic annihilation emission \cite{1990ApJ...358L..45S,1993ApJ...413L..85P}. Still, a few transient 511~keV flashes from X-ray transients have been reported, and remain a target to identify a unique high-energy source process, either pair plasma ejection, or short-lived $\beta^+$-decay radioactivity such as expected for novae (see next Section below).  

INTEGRAL has imaged the positron annihilation gamma-rays across the sky in great detail, and confirmed the diffuse nature of annihilation across our Galaxy \cite{2005A&A...441..513K}. The scientific surprise of this emission had already been apparent in earlier results from the Compton Observatory \cite{1997ApJ...491..725P}, and was consolidated by SPI measurements: The annihilation emission predominantly arises in the inner Galaxy in an extended region of size \about~10\degree. By comparison, the disk of the Galaxy is much fainter, wit a bulge-to-disk intensity ratio of 1.4 from a total luminosity of \about~2~10$^{-3}$ph~cm$^{-2}$s$^{-1}$ \cite{2005A&A...441..513K,2008Natur.451..159W}. 

\begin{SCfigure}
\centering
\includegraphics[width=0.68\textwidth]{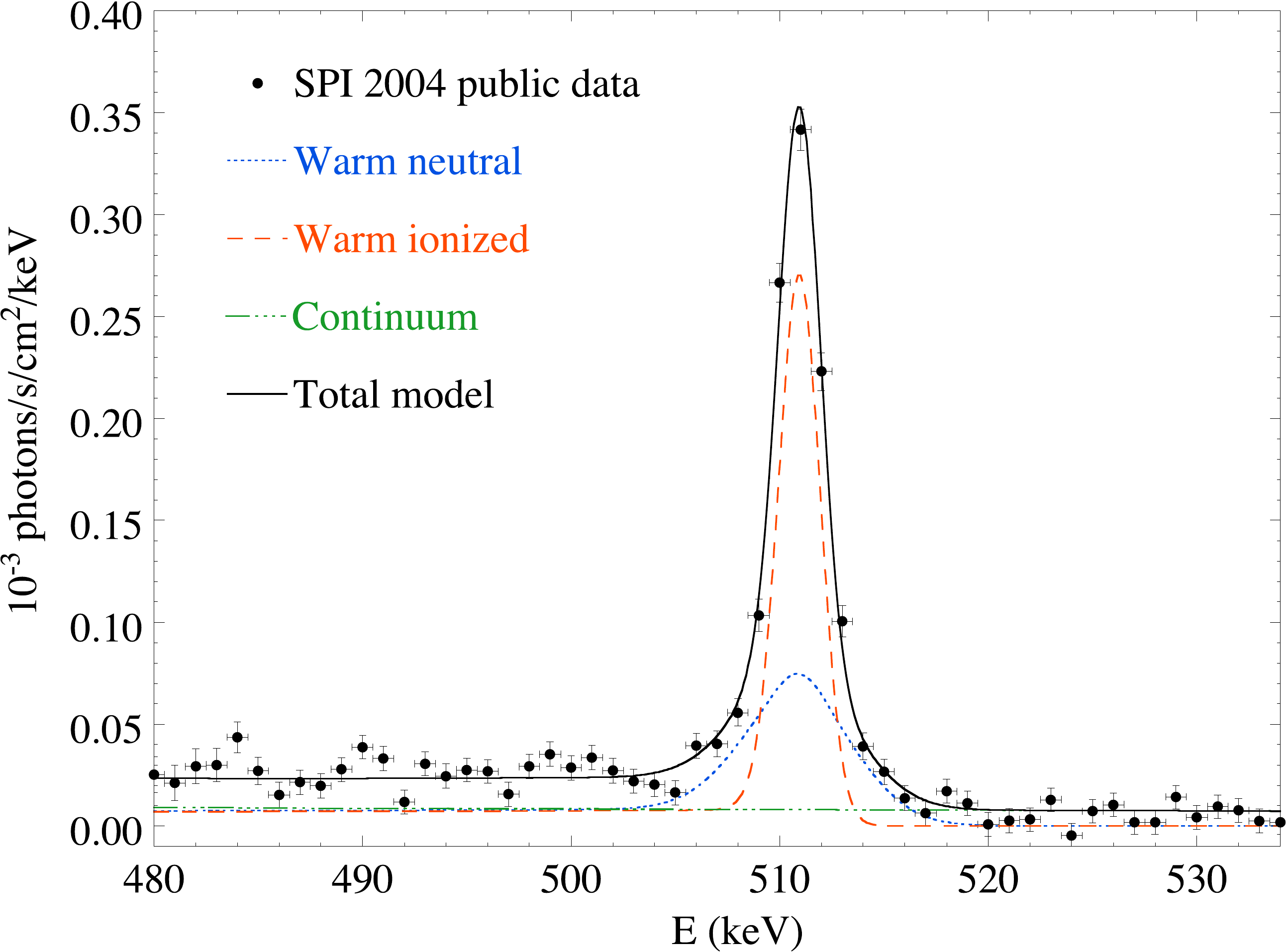}
\caption{The SPI spectrum of emission from positron annihilation in the Galaxy. The 511 keV line appears only moderately broadened, and, together with the high positronium fraction derived from the annihilation continuum $<$500~keV, supports annihilation to occur on average in moderately hot and only partially-ionized gas.}
\label{fig:511spectrum_SPI}
\end{SCfigure}

The line shape of the 511~keV line, and the line-to-continuum ratio of the emission from positron annihilation, both carry information about the conditions in the annihilation sites: Doppler broadening from the kinetic energy of the annihilating leptons is expected to increase the line width, unless annihilation occurs at low (thermal) energies only, or the momentum is transferred to a heavy atomic nucleus or dust grain surface where annihilation may occur. 

From the rather narrow 511~keV line, neutral gas appears to be an important ingredient of annihilation sites, and positrons form a positronium atom capturing an electron from interstellar hydrogen, before annihilating. On average, annihilation environments are only moderately ionized (percent level), and moderately hot at \about~8000~K; this suggests that annihilation does not occur in the hot interstellar medium that is characteristic for the surroundings of most candidate sources, but rather in the outer, partially-ionized boundary layers of molecular clouds. 

\begin{figure}
\centering
\includegraphics[width=0.48\textwidth]{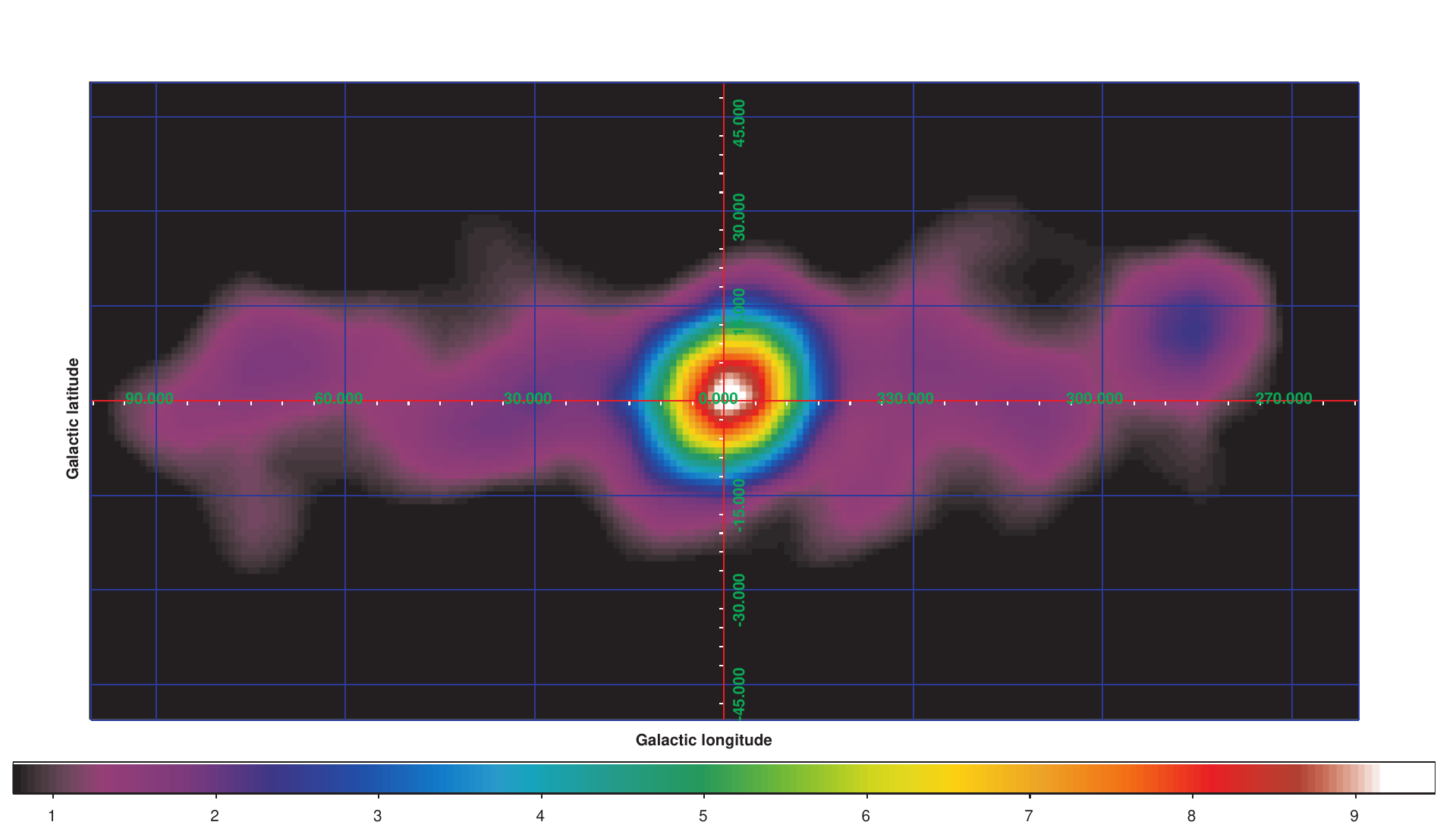}
\includegraphics[width=0.48\textwidth]{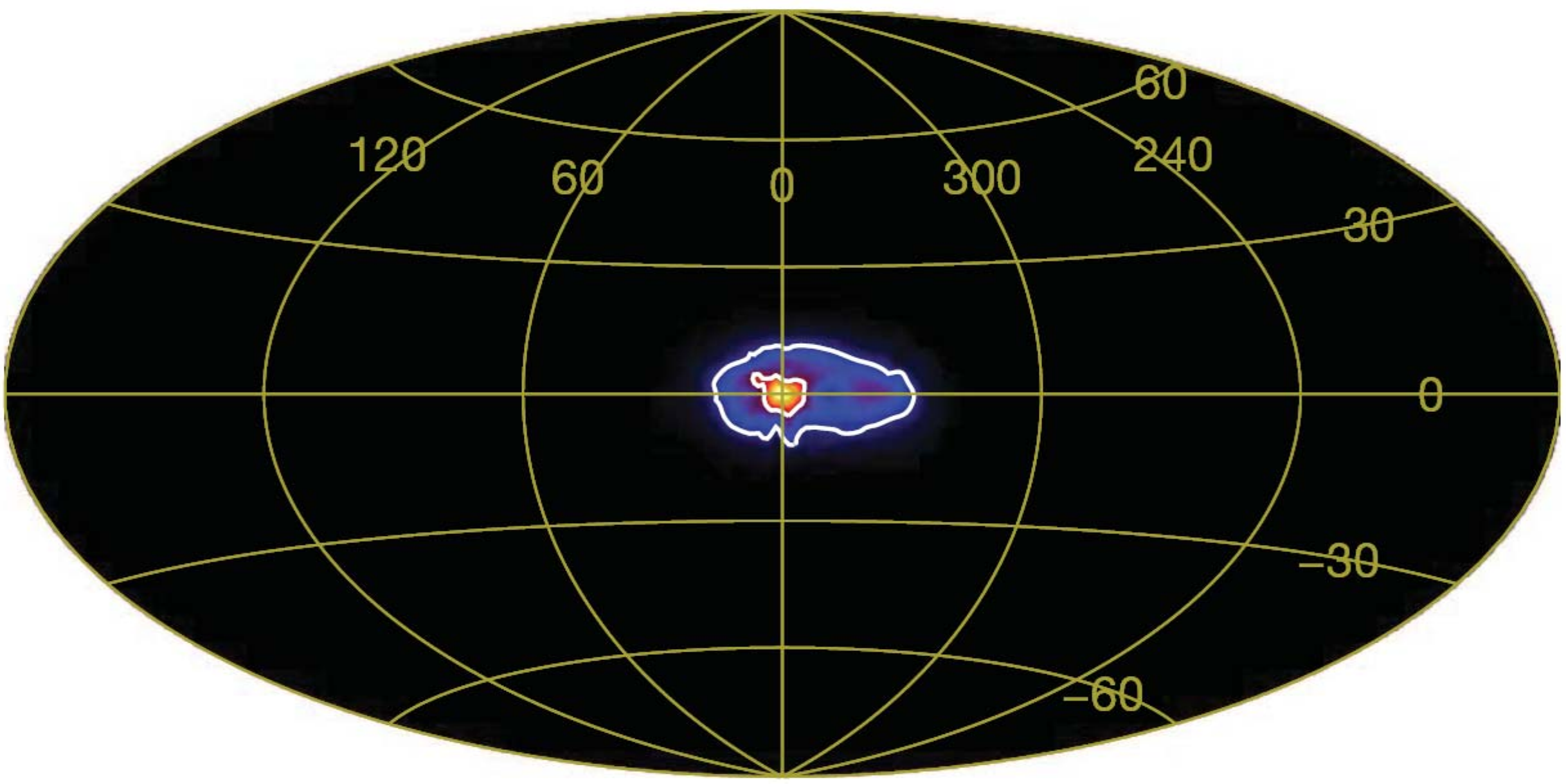}
\caption{The SPI images of Galactic annihilation emission show a dominating bright bulge-like (extended) emission centered in the Galaxy, with weak disk emission. \emph{left:} Significance map of a pixelized maximum-likelihood imaging deconvolution \cite{2010ApJ...720.1772B}.
\emph{right:} Intensity map of a multi-resolution expectation maximization imaging deconvolution \cite{2008Natur.451..159W}.
The centroid of the bulge emission may hold clues to the nature of the sources in the bulge, relating a small offset or an asymmetric inner-disk component (depending on analysis approach; see text) to spatial distributions as expected from stellar or, alternatively, dark-matter related sources.}
\label{fig:annihilation_image_SPI}
\end{figure}

In view of the variety of candidate sources of cosmic positrons \cite{2011RvMP...83.1001P}, it remains challenging to attribute positron source intensities to each of them. Although their positron injection energies may be different, and also their characteristic locations in the Galaxy may be different, positron propagation from their sources to their annihilation sites is quite uncertain. This leaves interpretational freedom for slowing down, and rearrangements of the spatial distributions. 
SPI images (figure~\ref{fig:annihilation_image_SPI}) include hints for deviations from the ideal symmetry expected for dark-matter origins of the bulge emission. Asymmetric disk emission was interpreted as possibly related to  X-ray binaries \cite{2008Natur.451..159W}. But in fact, each of the candidate sources may have its own deviations from perfect Galactic symmetry, and conclusions are model dependent and rely on the quality of the imaging information. Therefore, efforts in those studies focus on precise mapping of the annihilation emission and its correlation with source distribution models, and on searches from known / specific candidate positron-producing objects. 

\subsection{Novae}\label{sources_novae}
Nova explosions are understood as the nuclear ignition of the accreted surface layer  on a white dwarf. Nucleosynthesis is characterized by explosive hydrogen burning, producing preferentially nuclei on the proton-rich part of the valley of stable nuclei. This leads to the expectation of positron annihilation gamma-rays (see above) following the radioactive decay of those freshly-produced nuclei, with their often short lifetimes  \cite{1998ApJ...494..680J,2000MNRAS.319..350J}. 
However, this \emph{annihilation flash} would occur reveal days before the nova achieves its optical peak brightness, and thus would precede the detection and identification of the nova as a celestial source. Therefore, post-analyzing sky survey data for serendipitous exposures of such novae are needed to discover such a characteristic signature.
Recurrent novae probably only process small amounts of previously-accreted material, while classical novae accrete total amounts of 10$^{-4}$-10$^{-5}$~\Msol at very slow rate of order 10$^{-9}$~\Msol~y$^{-1}$ before ignition of the nova runaway.  
 
About 1/3 of all novae may be due to the more-massive and more-evolved white dwarf progenitors, which are enriched in heavier seed nuclei such as Na, Mg, and Ne. Here it appears plausible that H burning also passes through the Ne-Na cycle and leaves behind significant amounts of $^{22}$Na, a radioactive isotope with decay time of 3.8 years, and thus a candidate emitter of gamma-rays at 1274.53 and 511~keV. 
A major uncertainty in nova models is the amount of admixed white-dwarf material, another the total amount of ejected material. 
The variety of gamma-ray signals which could arise from novae of the CO and O-Ne types have been summarized in \cite{2004NewAR..48...35H}.

\begin{figure}
\centering
\includegraphics[width=0.78\textwidth]{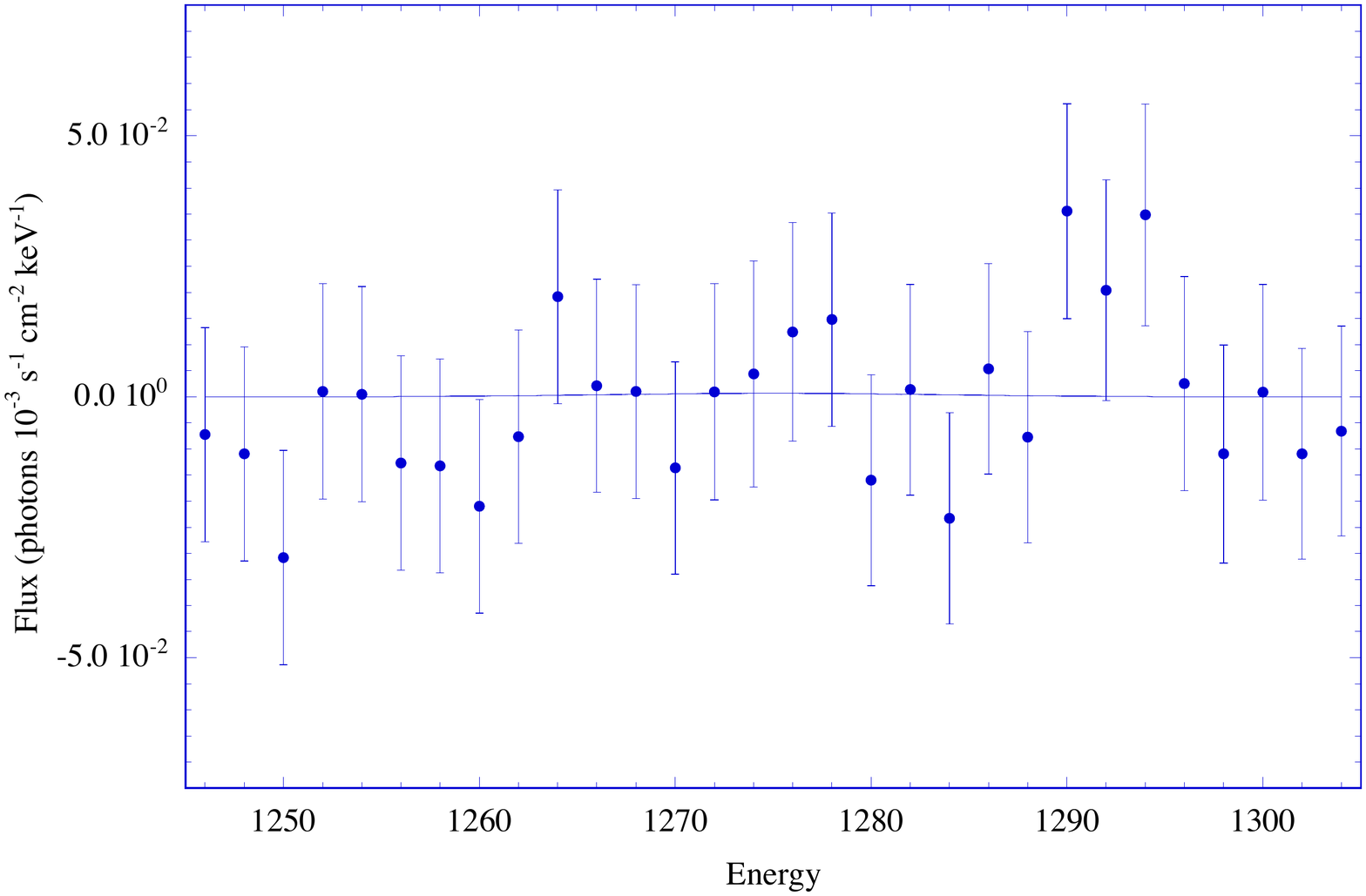}
\caption{SPI data from 3 years of data have been systematically searched  for a signal from the galactic distribution of novae \cite{2004ESASP.552..119J}. Here a spectrum shows the result in the vicinity of the $^{22}$Na line, derived testing a spatial source distribution as plausible for nova events \cite{1991ApJ...378..131K}. The fitted line at 1275~keV can hardly be seen, and corresponds to an (insignificant; 1$\sigma$) flux of 1.3~10$^{-5}$ph~cm$^{-2}$s$^{-1}$ (P. Jean, private communication).}
\label{fig:22Na_SPI}
\end{figure}

None of the gamma-ray missions so far  have resulted in a detection of any of the gamma-ray lines expected from novae, neither positron annihilation, nor $^{22}$Na nor $^7$Be decay signals have been seen. SMM \cite{Leising:1988}, WIND-TGRS \cite{Harris:2000a}, CGRO-COMPTEL \cite{Iyudin:1995,Iyudin:2001b}, and INTEGRAL \cite{Jean:2004} all have been used to search for these, and some hints in COMPTEL data seemed promising. But gamma-ray telescopes all suffer from intense activation and instrumental background in both the 511~keV and more so in the 1275~keV part of the measured spectrum from aluminum activation by cosmic ray bombardments in space.
As an example, we show the search for $^{22}$Na emission from novae with INTEGRAL/SPI \cite{2004ESASP.552..119J}. Here, the sky distribution was modeled from the Galactic nova distribution assumed to follow the 2.4~$\mu$m dust emission in the Galaxy \cite{1991ApJ...378..131K}, and was fitted to data from two years of  observations along the plane of the Galaxy, with 1.8~Ms of total exposure. Only an insignificant hint of the expected line at 1275 keV could be seen (figure~\ref{fig:22Na_SPI}). For the spatial model by  \cite{1991ApJ...378..131K}, which includes both a spheroidal and a significant disk population of novae, a 2$\sigma$ limit of 2.5~10$^{-4}$ph~cm$^{-2}$s$^{-1}$ was obtained. Translating this into a limit for the nova population attributed to this spatial distribution, adopting a nova rate of 20...40 per year and an O-Ne-Mg enriched nova fraction of 13...33\%, an average ejected-mass limit of 2.5--5.7~10$^{-7}$\Msol was obtained.

\subsection{Low-Energy Cosmic Ray Interactions}\label{diagnostics_cr}
Cosmic-ray interactions with interstellar gas are expected to produce spectral signatures at gamma-ray energies through excitation of nuclear levels, and through pion decay and higher-energy nucleonic excitations \cite{Ramaty:1979}. We concentrate on signatures from atomic nuclei here, leaving pion and GeV lines aside (see \cite{Strong:2007qe} for a review of continuum processes). In the nuclear line region, lines are expected from $^{12}$C and $^{16}$O at 4.43 and 6.1~MeV, respectively.
A nuclear line de-excitation spectrum including those C and O lines may be expected also from particle acceleration sites and sources, as energetic particles interact with ambient gas within or near the accelerating region. For a supernova remnant, a detailed prediction is found in \cite{2011A&A...533A..13S}, showing rather broad lines from C and O at 4.4 and 6.1~MeV. Estimated fluxes fall close to the sensitivity of COMPTEL;  due to the large line width, these are well below INTEGRAL's capabilities.

The capture of neutrons on hydrogen is expected to produce a line at 2.223~MeV, as thermalized neutrons form deuterium with hydrogen nuclei and liberate the neutron binging energy. Such gamma-ray line emission is prominent in instrumental lines, as atmospheric neutrons capture on hydrogen-rich propellants present in satellites \cite{Weidenspointner:2003}, but also could arise from the surface of compact stars (white dwarfs, neutron stars), in particular in accreting binary systems. Line shape information could be valuable, as orbiting material in accretion disks could be identified, or gravitational redshifts be exploited wrt. neutron star accretion geometry. A search for this line and corresponding point sources was negative, and only provided upper limits\cite{2004NuPhS.132..396G}. This is in line with expectations for the candidate sources and their plausible accretion rates.

Searches for above-discussed nuclear lines have been performed, from data addressing regions where cosmic ray interactions plausibly produce such excitations. Targets of interest are nearby clusters of massive stars, and along the inner ridge of the Galaxy where many of those star clusters should exist along any line of sight \cite{Bloemen:1997}. The measurements with high-resolution spectrometers are less sensitive to the typical and largely-broadened lines; therefore, INTEGRAL and RHESSI have not obtained useful measurements. But COMPTEL's scintillation detector resolution of \about~10\%~(FWHM) had been suitable for setting constraints. Unfortunately, the discussion and assessment of instrumental backgrounds had been difficult (and controversial) within the COMPTEL team, and conflicting results were obtained. An early report of nuclear lines from the Orion region \cite{Bloemen:1994a} created much excitement \cite{Bykov:1996}, also because the reported intensities would have implied very efficient cosmic-ray acceleration in this region \cite{Bykov:1994}. But after excluding possibly-contaminated parts of the data, the flux constraints had to be relaxed \cite{Bloemen:1999c}, and no clear line detection remained. It appeared that the line intensities expected from low-energy cosmic ray excitations both from Orion and from the inner Galaxy are just about near the instrumental limits of the COMPTEL observations. Thus, no new lessons can be taken yet, and must await a more-sensitive instrument/mission in the MeV energy band.  

\subsection{Our Sun and Solar Flares}\label{sources_sun}
Energetic particle acceleration is a key characteristic of solar flares, according to the established model \cite{2001ApJ...548..492S}: Reconnection events of magnetic field lines in the upper solar corona set up some kind of Fermi accelerator, at significant altitude above the solar chromosphere. It remains to be understood how the acceleration is set up and evolves \cite{2005AdSpR..35.1825M}. It seems clear that the energy for particle acceleration derives from the magnetic field, and that changes in its configuration set up the critical conditions to create relativistic particles. As a result, high-energy protons, nuclei, and electrons hit the deeper layers of the solar chromosphere, and produce a variety of observational signatures once interacting with denser gas towards the solar photosphere, at typical gas densities 10$^{14}$cm$^{-3}$ or higher. Among those signatures, electron Bremsstrahlung produces continuum emission, nuclear excitations produce a variety of lines, most prominently $^{20}$Ne, $^{12}$C, and $^{16}$O de-excitation at 1.634, 4.439, and 6.129~MeV, respectively, and spallation-produced neutrons capture on hydrogen and produce a characteristic line at 2.223~MeV. 

Most-detailed gamma-ray spectra had been measured for the 4 Jun 1991 solar flare with the OSSE spectrometer on CGRO, confirming these spectral characteristics \cite{1997ApJ...490..883M}. The modest spectral resolution of OSSE (\about~8\%~FWHM) implied that the most-valuable diagnostics were temporal changes of spectral-line intensities. For example, the relative intensities of the $^{20}$Ne and $^{16}$O lines encode the particle spectrum, because the nuclear-excitation cross sections for these isotopes are very different, $^{20}$Ne activation having a much lower excitation threshold energy. Analyzing this line ratio, and comparing with electron Bremsstrahlung, it was found that the charged-particle energy is about equally distributed between electrons and nuclei, and that the ion energy spectrum is steep with typical power-law slopes between -3.5 and -5.5 \cite{2005AdSpR..35.1825M}. 

\begin{SCfigure}
\centering
\includegraphics[width=0.6\textwidth]{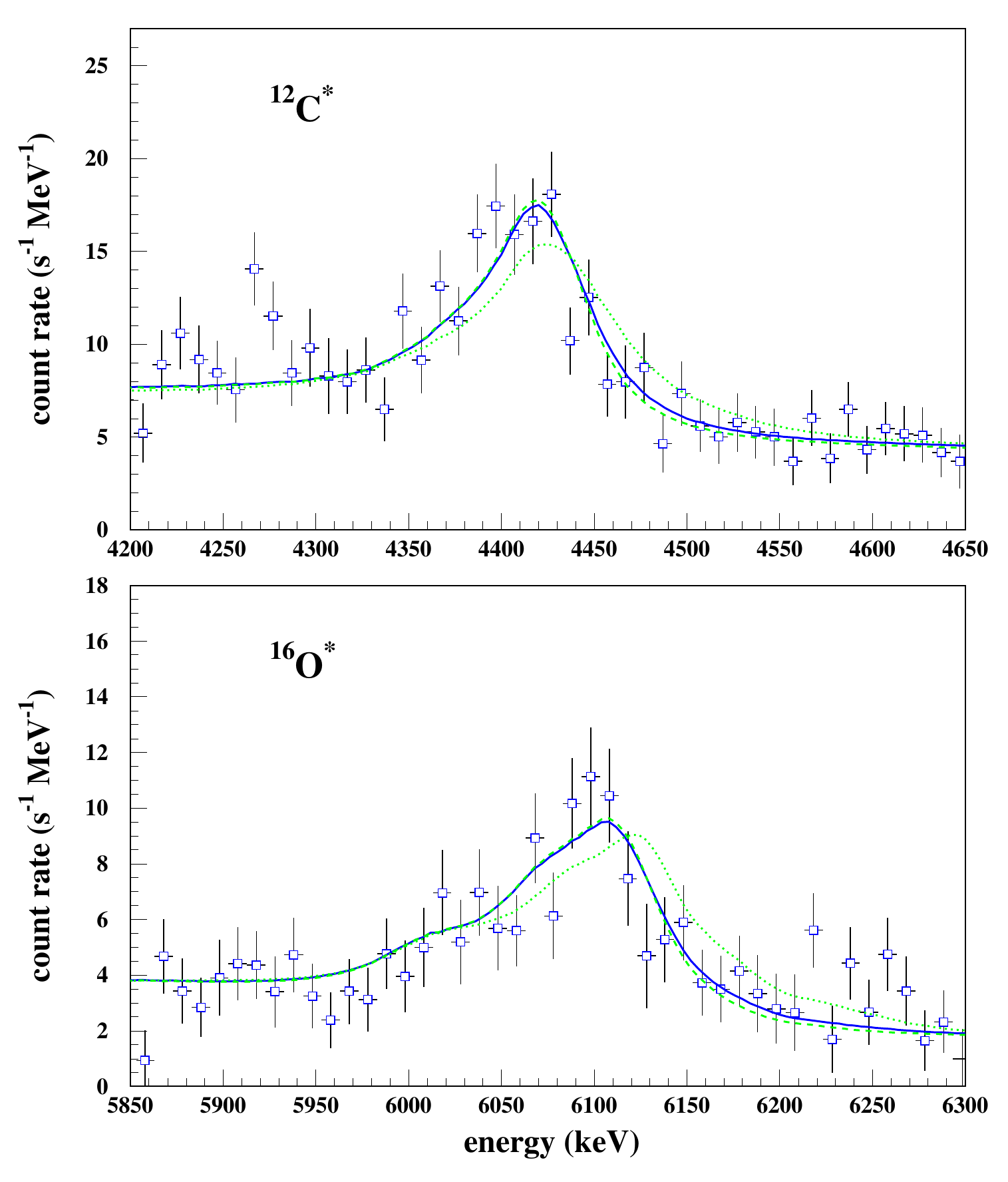}
\caption{The SPI measurements of the de-excitation emission from $^{12}$C \emph{(above)} and $^{16}$O
\emph{(below)} reveal low-energy tails. These arise from pitch-angle scattering of solar-flare particles, as they hit the solar atmosphere from above. See \cite{2006A&A...445..725K}  for details.}
\label{fig:SolFlare_C_O}
\end{SCfigure}

Important diagnostics from high-resolution spectroscopy employs Doppler shifts of gamma-ray lines or attenuation between source and observer, which modify the position and shape of the gamma-ray lines (Fig.~\ref{fig:SolFlare_C_O}). Flare-accelerated particles propagate downward, spiraling around magnetic field lines. Depending on the importance of scattering, and on convergence of field lines, particles may propagate deeper into the chromosphere, or get magnetically mirrored at higher altitudes when pitch angles are large. Therefore the line shape details encode how accelerated particles lose their energy after the acceleration event and before thermalization.

The \emph{Ramaty High-Energy Solar Spectrometer Instrument (RHESSI)} \cite{2002SoPh..210....3L}  included a gamma-ray spectrometer with 9 Ge detectors,  located behind a rotation-modulation collimator for high-resolution imaging at 2.3~arcsec resolution. RHESSI always points to the Sun. Solar-flare data collected with SPI/INTEGRAL typically occur with the instrument pointed away from the Sun, and gamma-rays entering Ge detectors through the side and penetrating the BGO scintillation detector which forms SPI's anti-coincidence system.

A bright flare occurred on 28 Oct 2003, early in both those missions, and resulted in bright gamma-ray line signals extending over a period of \about~15 min \cite{2006A&A...445..725K}. 
The gamma-ray flare started off with intense continuum emission, but within \about~a minute, line emission set in, with nuclear de-excitation lines preceding the neutron capture line at 2.223 MeV, as expected from the required slowing-down processes.
Nuclear lines were identified from neutron capture on hydrogen, and from excited $^{12}$C, $^{16}$O, $^{24}$Mg, $^{20}$Ne, and $^{28}$Si nuclei. The line intensity ratios evolve somewhat over the period of the flare. In particular the C and O line ratios are sensitive to the ratio of $\alpha$ particles over protons in the flare, and suggest that the $\alpha$/p ratio also evolves during the flare itself, reducing the $\alpha$-particle content in the later flare phase.
SPI data from these lines  detected of kinematic Doppler shifts/distortions (see Fig.~\ref{fig:SolFlare_C_O}).  These were interpreted as a beam geometry of accelerated particles being downward-directed and rather fan-like, which implies that pitch-angle scattering in the solar atmosphere is large. 

Since mid 2010, again more solar flare events are being observed as solar activity begins to increase towards the next solar maximum in mid 2013. Observations now also include measurements from the GBM scintillation detectors aboard the Fermi satellite. Studies focus on discrimination of different energy ranges, as they determine nuclear lines, the 2.23 MeV neutron capture line, and the spectral signature related to pions (in the order of increasing accelerated-particle energies), so try and learn about the energetics of the solar-flare particle acceleration process.

\subsection{Nuclear Absorption of Background Gamma-Rays}\label{diagnostics_abslines}
Most of the spectroscopic signatures which are exploited in astronomy are related to absorption processes, which shape a continuum background light source in ways characteristic for the absorbing atoms or molecules.  Atomic nuclei should, by analogy, provide absorption signatures which are characteristic for nuclei. This will be particularly useful where matter is fully ionized, such as expected in the vicinity of gamma-ray burst sources or active galactic nuclei. Both these sources generate intrinsic photon emission with a featureless continuum spectrum extending to gamma-ray energies.  

Candidate nuclear absorption from early-universe cosmic matter could arise from nuclear excitation of abundant nuclei such as C and O (see above), but probably more significantly from He excitation and dissociation through the giant dipole resonance with a threshold energy of \about~25~MeV.   High redshifts of the most-interesting objects move such features into the \about~MeV regime. Gamma-ray telescopes so far could not find these signals, which should, however, be detectable with next-generation gamma-ray telescopes \cite{2011ExA...tmp..116G}. 

\section{Summary and Conclusions}
The pioneering balloone-borne experiments (\about~1976) and the  COMPTEL \cite{Schoenfelder:1993a} and OSSE \cite{Johnson:1993} instruments aboard the Compton Observatory (1991-2000; \cite{Gehrels:1993}) provided a first sky survey and identified several prominent gamma-ray lines, thus confirming expectations from theory. The INTEGRAL mission \cite{Winkler:2003} deepened the exploration of the nuclear-radiation sky since 2002. Gamma-ray spectroscopy was thus consolidated with the measurements at sufficient spectral resolution of Ge detectors through ESA's INTEGRAL mission and its spectrometer SPI \cite{Vedrenne:2003}. The sensitivity and astronomical resolution of all these instruments is comparable (fluxes must be above 10$^{-5}$ph~cm$^{-2}$s$^{-1}$ and sources can only be resolved on the degree scale), and only the brightest sources can be seen due to backgrounds intrinsic to all instruments.
The predicted and known gamma-ray lines from radioactive isotopes have been confirmed, measurements have been enriched with spectroscopic detail which add significant astrophysical aspects. 

The characteristics of  $^{44}$Ti emission from the Cas A supernova remnant and candidate sources in the Galaxy  show that  nucleosynthesis in core-collapse supernovae occurs under a diversity of environmental conditions, which result from  substantial deviations from spherical symmetry in these explosions \cite{The:2006}. 
Exploitation of gamma-rays from the $^{56}$Ni decay chain which is responsible for supernova light still awaits a sufficiently-nearby supernova of type Ia \cite{Isern:2013}; the independent information carried by gamma-rays from this decay chain could help to understand details of explosion physics, which happens in the dense and dynamic initial phase of the exploding star that is otherwise occulted to observations.
Diffuse gamma-ray emission from $^{26}$Al decay \cite{Diehl:2011} as well as positron annihilation gamma-rays at 511~keV \cite{Prantzos:2011} have established their own astronomical windows, where unique measurements are being made.  The detailed multi-messenger comparison of their emission in intensity and spatial morphology with model predictions for each specific candidate sources provides insights into massive-star groups and high-energy sources throughout the Galaxy. The puzzle of the bright bulge-like 511~keV image holds a promise for learning new aspects both of cosmic-ray propagation and of dark-matter interactions in our Galaxy. 
$^{60}$Fe radioactivity from (probably) the same sources which create \Al provide a global massive-star population diagnostic of the interior structure of massive stars. The discovery of $^{60}$Fe also in  ocean crust material \cite{Knie:2004} underlines the role of such long-lived radioactive isotopes as tools to study  cosmic nucleosynthesis, in particular also addressing the aspect of how core-collapse supernovae spread their ejecta in their surroundings. \Al emission from nearby regions is bright enough so it can be located relative to its sources. 
Fine spectroscopy of the \Al line in the inner Galaxy shows Doppler shifts from the large-scale motion of galactic rotation with surprisingly-high velocities. This may teach us new aspects through nucleosynthesis ejecta of how massive stars shape their surroundings and provide feedback that regulates star formation in disk galaxies.
Positron annihilation throughout the Galaxy shows puzzling brightness of the bulge region of our Galaxy, which may find explanations in the complex propagation of positrons in tangled, but possibly also rather regular magnetic-field configurations in interstellar space of the Galaxy's disk and halo. It may also shed light on past activity in the center of our Galaxy, or even interactions of dark matter.
The acceleration of particles to the relativistic energies of cosmic-rays are still unclear a hundred years after cosmic rays had been discovered \cite{Blasi:2013}. How thermal particles develop high-energy tails, how these are then boosted towards relativistic energies,  and where this happens, this remains to be one of the big astrophysical questions. Gamma-ray spectroscopy can address the low-energy part of physical processes herein, and detailed spectra of these same lines are being studied from solar flares and their gamma-ray spectra. 

Nuclear line emission is expected from more and different sources than the ones known so far, such as nova explosions or the surfaces of compact stars in binary systems, and from other isotopes than the ones discovered so far. The unique and different processes which generate nuclear line features are only partly exploited with respect to their astrophysical messages. An advance in sensitivity is needed, to reach other nearby galaxies in the known lines, and to find several more of the expected. Ideas for such next-generation telescopes have been presented, but not given priority by evaluating committees. Exploitation of the current instruments and missions can still add important detail, on the Galactic sources, and on instrument building and data analysis approaches. The nuclear-physics signatures of cosmic processes can teach us aspects that are difficult to reveal by other means.

\noindent{\bf Acknowledgements.}
The INTEGRAL/SPI project has been completed under the responsibility and leadership of CNES; we are grateful to ASI, CEA, CNES, DLR (grants 50 OG 1101 and 50 OR 0901), ESA, INTA, NASA and OSTC for support of this ESA space science mission and its science analysis. 
This research was supported also by the German DFG cluster of excellence \emph{Origin and Structure of the Universe}. 

\bibliographystyle{abbrv}

\end{document}